\documentclass[a4paper,fleqn,usenatbib]{mnras}

\usepackage[T1]{fontenc}
\usepackage{ae,aecompl}

\usepackage{graphicx}	
\usepackage{amsmath}	
\usepackage{amssymb}	
\usepackage{wasysym}
\usepackage{makecell}

\title[Assessing retention fractions for BH-BH mergers in GCs]{MOCCA-SURVEY Database I: Assessing GW kick retention fractions for BH-BH mergers in globular clusters}

\author[J. Morawski et al.]{
Jakub Morawski,$^{1}$
M. Giersz,$^{2}$
A. Askar,$^{2,3}$
K. Belczynski$^{2}$
\\
$^{1}$Astronomical Observatory, Warsaw University, Al. Ujazdowskie 4,  Warsaw 00-478 Poland\\
$^{2}$Nicolaus Copernicus Astronomical Center, Polish Academy of Sciences, ul. Bartycka 18, Warsaw 00-716 Poland\\
$^{3}$Lund Observatory, Department of Astronomy, and Theoretical Physics, Lund University, Box 43, SE-221 00 Lund, Sweden
}

\date{Accepted XXX. Received YYY; in original form ZZZ}

\pubyear{2017}

\begin{document}
\label{firstpage}
\pagerange{\pageref{firstpage}--\pageref{lastpage}}
\maketitle

\begin{abstract}
Anisotropy of gravitational wave (GW) emission results in a net momentum gained by the black hole (BH) merger product, leading to a recoil velocity up to $\sim10^3\text{ km s}^{-1}$, which may kick it out of a globular cluster (GC). We estimate GW kick retention fractions of merger products assuming different models for BH spin magnitude and orientation (MS0 - random, MS1 - spin as a function of mass and metalicity, MS2 - constant value of $0.5$). We check how they depend on BH-BH merger time and properties of the cluster. We analyze the implications of GW kick retention fractions on intermediate massive BH (IMBH) formation by repeated mergers in a GC. We also calculate final spin of the merger product, and investigate how it correlates with effective spin of the binary. We used data from MOCCA (MOnte Carlo Cluster simulAtor) GC simulations to get a realistic sample of BH-BH mergers, assigned each BH spin value according to a studied model, and calculated recoil velocity and final spin based on most recent theoretical formulas. We discovered that for physically motivated models, GW kick retention fractions are about $30\%$ and display small dependence on assumptions about spin, but are much more prone to cluster properties. In particular, we discovered a strong dependence of GW kick retention fractions on cluster density. We also show that GW kick retention fractions are high in final life stages of the cluster, but low at the beginning. Finally, we derive formulas connecting final spin with effective spin for primordial binaries, and with maximal effective spin for dynamical binaries.     
\end{abstract}




\section{Introduction}



\begin{figure}
\centering
\includegraphics[width=\columnwidth]{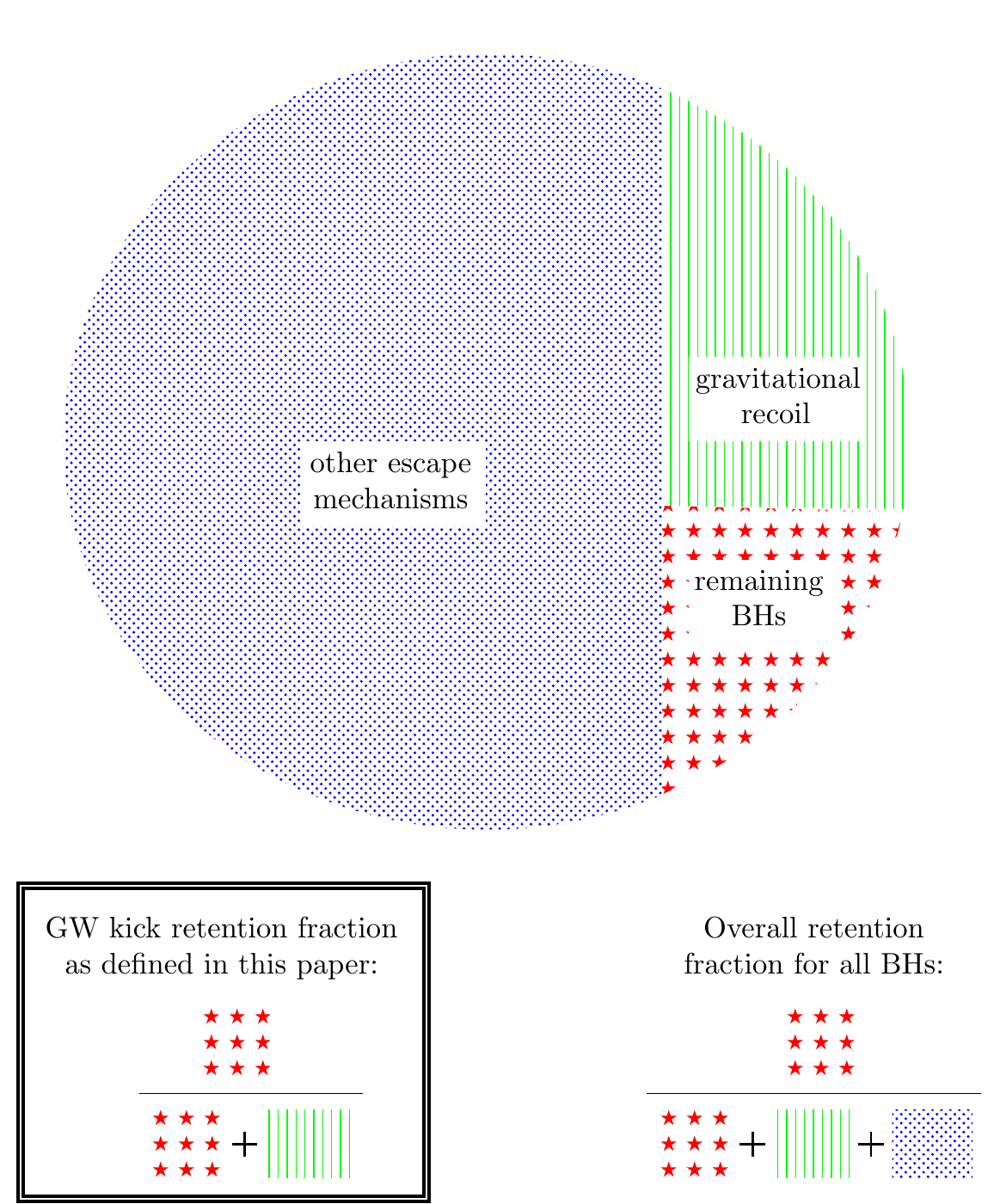}
\caption{Chart illustrating how the GW kick retention fraction is defined for the sake of this paper, as opposed to an overall retention fraction for all BHs. We only consider these BHs which have not been ejected due to other processes.}\label{whatisafraction}
\end{figure}

There are several processes which can lead to BH ejection from a GC, including natal kicks, dynamical interactions, escapers due to relaxation process and tidal influence of a host galaxy. Some BH-BH binaries, which are retained inside of the cluster (about $10\%$ of all such binaries, see \citet{MOCCA2}) and eventually merge into a single BH subsequently can be ejected due to a GW kick experienced during the merger. When two BHs merge and emit a gravitational wave, the process is not perfectly symmetrical, with an effect of a non-zero net momentum carried away with the wave (phenomenon first analyzed in \citet{Firstcalculationsofgravitationalrecoil}). To preserve the conservation of momentum law, the new BH created due to the merger receives a kick in the opposite direction. Such a kick can be very strong, up to the order of magnitude of $10^3\text{ km s}^{-1}$, leading to an escape from a GC. This process is the focus of our study, therefore by GW kick retention fraction we refer to a fraction of BH-BH mergers occurring inside a cluster which are retained despite the GW kick, not the overall retention fraction for all BHs and all escape processes. We explain this in Fig. \ref{whatisafraction}\\
If GW kick retention fractions are low, it would hinder formation of IMBHs (for the sake of this paper by term IMBH we will refer to any BH above $100M_{\astrosun}$ and below $10^5M_{\astrosun}$), as it blocks one of the formation channels: consecutive gravitational bonding and merging with other BHs in a cluster (\citet{MillerHamilton}, \citet{MOCCA2015}, \citet{MOCCA2}). Since the existence of IMBHs is still essentially hypothetical, but with LIGO and LISA gravitational wave observations we should soon accumulate data on the whole range of BH masses, it would be very valuable to make some predictions concerning the probability of IMBH occurrence. Assessing typical kick magnitudes, GW kick retention fractions and merger product spins plays an important role in understanding how BH-BH binaries in clusters evolve, therefore our results will impact the formation of IMBHs and/or BH subsystems and the way how a cluster evolves, which makes it crucial to incorporate them into future GC simulations (including these realistic recoil velocities and final spins in the Monte Carlo MOCCA code (\citet{MOCCAHypki}, \citet{MOCCA}) will be our next goal). With future observational data it will be possible to test and calibrate such results while also learning more about BH spins. In this paper we present results of modelling spinning BH-BH binaries with various assumptions about correlation
between natal spin magnitude and mass, and about spatial orientation of the two spin vectors.\\
The question of retention probability for BH-BH mergers due to GW kicks has been previously addressed in several publications. \citet{Meritt} analyzed lower and upper bounds on GW kick magnitudes based on approximate formulas known at the time and compared them with typical escape velocities in different types of galaxies and GCs, reaching the conclusion that a BH-BH merger product is very unlikely to be retained in a GC. \citet{OLeary} performed semi-analytical  simulations to assess maximal mass of a retained BH in a GC, reaching the conclusion that probabilities of harbouring an IMBH in a GC are very low, but increase with core density. \citet{MBH} derived retention fractions for massive black holes (any masses above $100M_{\astrosun}$) in galaxies assuming certain mass-ratio and spin distributions, obtaining high retention probability especially in older galaxies. \citet{IMBHwithBH} assessed, on the bases of semi-analytical models, the probability of retention of an IMBH in a GC after repeated mergers (typically 25 mergers with less massive BHs), considering several models for spin magnitude, concluding that non-zero ejection probability virtually always exists, but IMBHs with $M\gtrsim1000M_{\astrosun}$ are very likely to be retained. \citet{NSclusters} analysed BH-BH mergers in nuclear star cluster, showing that contrary to typical GCs their GW kick retention fractions  should be high, thanks to higher escape velocities. \citet{retentionMoody} derived GW kick retention fractions in different types of GCs assuming zero-spin formula, concluding that highest retention probability occurs in massive, metal-rich clusters (reaching almost $50\%$ while $10-25\%$ GW kick retention fractions were found in less favourable clusters). \citet{AmaroSeoane2016} used semi-analytical models to estimate mass ratios, spin magnitudes and other properties in BH-BH relativistic mergers.\\
Most of the aforementioned papers were focused on searching for potential LIGO sources and treated the issue of merger product retention probability as a side study. GW kick retention fractions have so far mostly been estimated based on semi-analytical methods involving theoretical distributions for binary properties, usually using simplified assumptions about spins or neglecting them altogether, and often were limited to a specific subclass of mergers. Work presented here is the first step in developing an alternative method, which is to use a sample of BH-BH binaries inherent to an advanced GC simulation (in our case MOCCA), and to track spin magnitudes for BHs entering repeated mergers (see Section \ref{trackingblackholes}). We also try to be more careful with initial assumptions about spins by adapting two different prescriptions for orientation depending whether a merger is dynamical or primordial (see Section \ref{Spin orientation}). Finally we study a new model for spin magnitude dependent on mass and metallicity introduced in \citet{Massspinrelation}. Our main goal is to develop the technique of determining GW kick retention fractions, but the secondary goal is to study this new model for spin magnitudes in the context of GW kick retention and see if resulting fractions differ significantly from those derived under simpler assumptions. \\
In Section \ref{About data} we will describe the simulation data we used for our study. In Section \ref{trackingblackholes} we discuss how we keep track of merging BHs and their subsequent spin values. In Section \ref{Approximations primordial and dynamics} we will introduce two classes of mergers (primordial and dynamical) for which in our study we assumed different properties regarding spin orientation.  In Section \ref{Allaboutkicks} we will describe the nature of gravitational recoil and provide relevant formulas. In Section \ref{Finalspin and effective spin} we will provide formulas for the effective and actual spin of the merger product. In Section \ref{Spin distributions} we will define our assumptions and models for initial BH spin magnitude and orientation. In Section \ref{Finding rates} we will describe our main result about the retention fractions. In Section \ref{Finalspin vs effective spin} we will analyse correlations between parameters measuring merger product spin introduced in Section \ref{Finalspin and effective spin}. In Appendix \ref{crossindetail} we will extend the analysis from Section \ref{Finalspin vs effective spin}.   
\section{Model}\label{About data}
We used data from MOCCA simulations: Monte Carlo cluster simulations which take into account both the relaxation process responsible for the dynamical evolution of the system, stellar evolution of single and binary stars (BSE code, \citet{BSE}), direct integration for few body interactions based on code of \citet{NbodyforMOCCA}, and realistic treatment of escape processes based on \citet{TreatmentofescapeprocessesinMOCCA}. For complete description of MOCCA refer to \citet{MOCCAHypki}, \citet{MOCCA}. \\
We used data from about 2000 simulations with various initial parameters, defining cluster properties, star and binary population, etc. (for details refer to \citet{MOCCA2015}; \citet{MOCCA2}). In our study we did not differentiate between these parameters, except for the initial cluster concentration, which is a proxy for the initial cluster density.\\
It is important to point out that we do not evaluate mergers which occurred outside of the cluster after a binary escaped from a GC - this is a subject for a different study. Our sample of BH-BH mergers consists of all cases found in MOCCA simulation outputs when a binary merged inside a cluster due to gravitational wave emission. We do not make any restrictions on BH mass nor merger time. There are about 4450 such cases. For these we extract from MOCCA several parameters, most importantly: masses of binary components, merger time, cluster escape velocity at the time of merger, metallicity of the model, and initial cluster global parameters. Our default treatment for assigning spin magnitudes is to calculate them according to a specific model for mass-spin relationship. However we only apply this for BHs that have not merged before in a given MOCCA simulation. While analyzing data from the same simulation we keep track of individual BHs by their identifier numbers (unique id's), so that if a BH entering a merger is already a product of a previous merger (from now on we will be referring to these as post-merger BHs), we apply a different method, described in Section \ref{trackingblackholes}. After spin vectors are assigned, we evaluate kick velocity from formulas described in Section \ref{Allaboutkicks}. \\ 
It is important to point out that the properties of compact binaries outputted by MOCCA strongly depend on the treatment of the common envelope phase (CEP) in binaries. CEP is an important phase in the evolution of interacting binaries. Despite numerous studies of the subject the energy transfer during CEP is not understood very well. It is usually described using two parameters: $\alpha$ and $\lambda$. $\alpha$ is the ratio between energy used to unbind the envelope and the difference in orbital energy $\Delta E_{\text{orb}}$ (before and after the CEP):
\begin{equation}\label{alphaCEP}
E_{\text{b,CEP}}=\alpha\Delta E_{\text{orb}}
\end{equation}
where $E_{\text{b,CEP}}$ is donor binding energy at CEP. It is possible to have $\alpha>1$, because there are other energy sources besides orbital energy, such as reionization energy, enthalpy (e. g. \citet{CEPEnthalpy}), or accretion energy if inspiralling object is a neutron star (NS) or BH. $\lambda$ is the binding energy parameter, which depends on the structure of the primary star. Equation \ref{lambdaCEP} expresses $E_{\text{b}}$, binding energy of the star, in terms of $\lambda$, stellar mass $M$, star core mass $M_{\text{c}}$ and stellar radius $R$. 
\begin{equation}\label{lambdaCEP}
E_{\text{b}}=\frac{GM(M-M_{\text{c}})}{R\lambda}
\end{equation}
For a more detailed explanation of the CEP and $\alpha$ and $\lambda$ parameters refer to \citet{AlphalambdaCEP}, \citet{CEPStarTrack}. The version of the BSE code (\citet{BSE}, \citet{BSE2}) we used assumes for the parameters describing the common envelope phase: $\alpha=3$, $\lambda=0.5$. These parameters are standard for the code. As a result a lot of binaries survive to the point where both components collapse into BHs, and a majority of mergers are primordial (without any strong dynamical interactions). \\
Recent studies suggest that these values for $\alpha$ and $\lambda$ might be wrong. In \citet{CEPStarTrack} it is shown that $\lambda$ is generally low for NS and BH progenitors, usually not exceeding $0.2$. \citet{Giacobbo} and \citet{Mapelli} consider different vaues for $\alpha$ and argue that $\alpha\approx5$ can be a reasonable value for NS and BH progenitors. Therefore it might be necessary to reassert our assumptions regarding the CEP in the future - different $\alpha$ and $\lambda$ values will result in different properties of MOCCA BH-BH mergers, which will in turn affect GW kick retention fractions.\\
\section{Tracking BHs}\label{trackingblackholes}
While analyzing data from the same MOCCA simulation we kept track of merger products and their spins. If a BH entered a merger for the second time instead of assuming spin in accordance with a given mass-spin formula, we would assume the final spin $a_{\text{fin}}$ (equation \ref{afin}) of the previous merger product. Also we did not evaluate mergers which should not occur because they included a post-merger BH which was not retained due to gravitational recoil. As a result, our sample is reduced from about 4450 cases down to about 3750 cases, depending on the studied model of mass-spin relationship. At the current stage no better approach is available, because previous dynamical studies of GC evolution (including MOCCA) have not accounted for kicks due to GW merger of two BHs. Unfortunately this method certainly is not fully consistent - we remove some BHs which are not removed in the actual MOCCA simulation, and they impact the evolution of the cluster.\\
We can only speculate how the GW kick retention fraction and the evolution of a cluster will be affected by removing these additional BHs. According to \citet{IMBHslow} removing a high mass BH will in consequence lead to contraction of the BH subsystem to increase the energy generation rate required to support the overall cluster structure (\citet{gravothermaloscillations}). The larger number of dynamical interactions probably will lead to minor increase of BH mergers. Involvement of lower and lower mass BHs will lead to larger GW kicks if there is any correlation between BH spin and the initial stellar mass (e.g. \citet{Massspinrelation}). In turn the retention fraction will be smaller. Removing more masssive BHs could hamper the formation of IMBH (if such were to form). This will also increase the number of BH-BH mergers and reduce the GW kick retention fraction (if the IMBH were to form it would quickly clear out the environment of other high mass BHs, which could otherwise merge with each other). An additional effect of tracking is that post-merger BHs entering another merger usually have higher spins, and this in turn leads to higher GW kicks and reduces the probability of retention. Concluding, on the basis of this raw analysis our results should be seen as estimations, most likely upper bounds on GW kick retention fractions.
\section{Primordial and dynamical mergers}\label{Approximations primordial and dynamics}
Stellar dynamics discerns two significantly different classes of binaries: primordial and dynamical binaries. The first class consists of all cases in which a binary is formed by two stars which were born together and belonged to the same binary for all their evolution history. Therefore by primordial mergers we refer to BH-BH binaries formed this way, with no exchanges at any evolution stage up to the merging point. Binaries formed at a later stage from stars which had individual evolution histories before they became a binary (due to exchanges and other interactions), are called dynamical binaries, and BH-BH mergers in such binaries will also be referred to as dynamical.\\
It is important to emphasize that neither the BHs belonging to a primordial binary or a primordial merger product should be confused with primordial BHs formed at the beginning of the Universe (\citet{primordialblackholes}), this is merely an unfortunate coincidence of terms which mean different things in different areas of astronomy.\\
In our data from MOCCA there were about 1050 dynamical mergers, but due to tracking the sample was reduced to about 350 dynamical mergers, depending on the studied model ($\sim10\%$ of all considered cases).\\
These two classes of BH-BH binaries have different properties.  There are two models for initial properties of primordial binaries. For 95\% initial binary fraction models, MOCCA draws binary component masses according to \citet{Kroupa} (see Table 1 in \citet{MOCCA2}). This means that for stars with mass $M>5M_{\astrosun}$ binaries are coupled by order pairing, i. e. to keep $q$ as close to $1$ as possible. In other cases a standard model of flat $q$ is applied, so it can be much lower.  Nevertheless for BH-BH mergers the first case is more common (because of higher binary fraction), thus most primordial binaries have $q$ very close to $1$. In addition since components of primordial binaries were not able to grow in mass by mergers, the masses of primordial binary components are usually low. Most dynamical mergers in our sample have very low $q$. In fact, in a majority of dynamical cases (almost $80\%$ of evaluated cases) one of the components is an IMBH, leading to mass ratio $q\approx10^{-3}-10^{-1}$.\\      
\section{Gravitational recoil}\label{Allaboutkicks}
\begin{figure}
\centering
\includegraphics[width=\columnwidth]{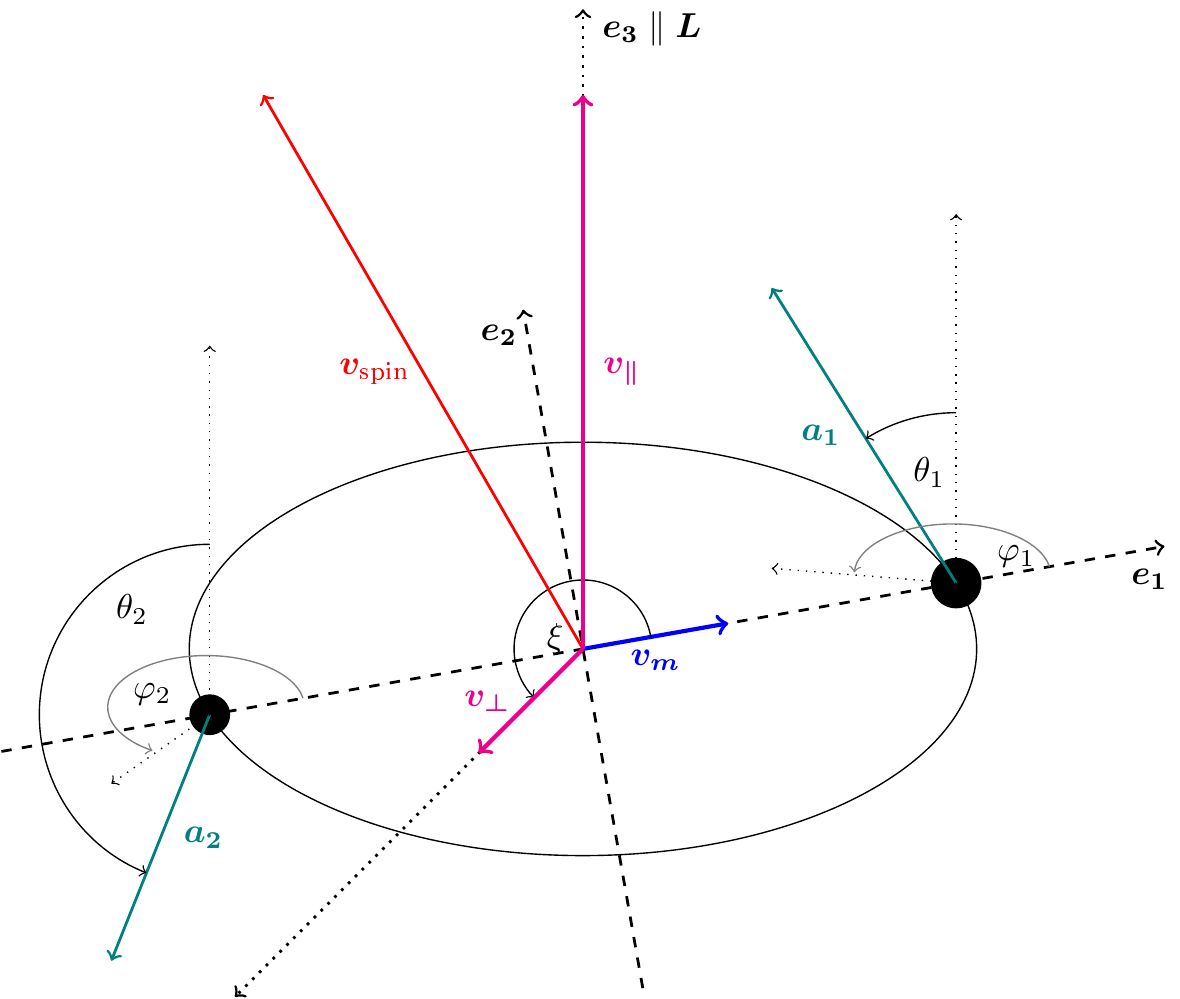}
\caption{Conceptual picture presenting components of the recoil velocity. Dashed lines represent a Cartesian coordinate system in the orbital plane: $\boldsymbol{e_1}$, $\boldsymbol{e_2}$. A vertical dotted line is a line perpendicular to orbital plane ($\boldsymbol{e_3}$, parallel to orbital angular momentum). Red vector is the kick component related to spin asymmetry, and magenta vectors are it's projections on the plane and parallel to $\boldsymbol{e_3}$. Blue vector represents the mass inequality contribution. Black filled circles represent a pair of BHs, their spins and orientation in a spherical coordinate system is illustrated. This drawing also reflects typical proportions between recoil velocity components.}\label{Kick components drawing}
\end{figure}

\begin{figure}
\centering
\includegraphics[width=\columnwidth]{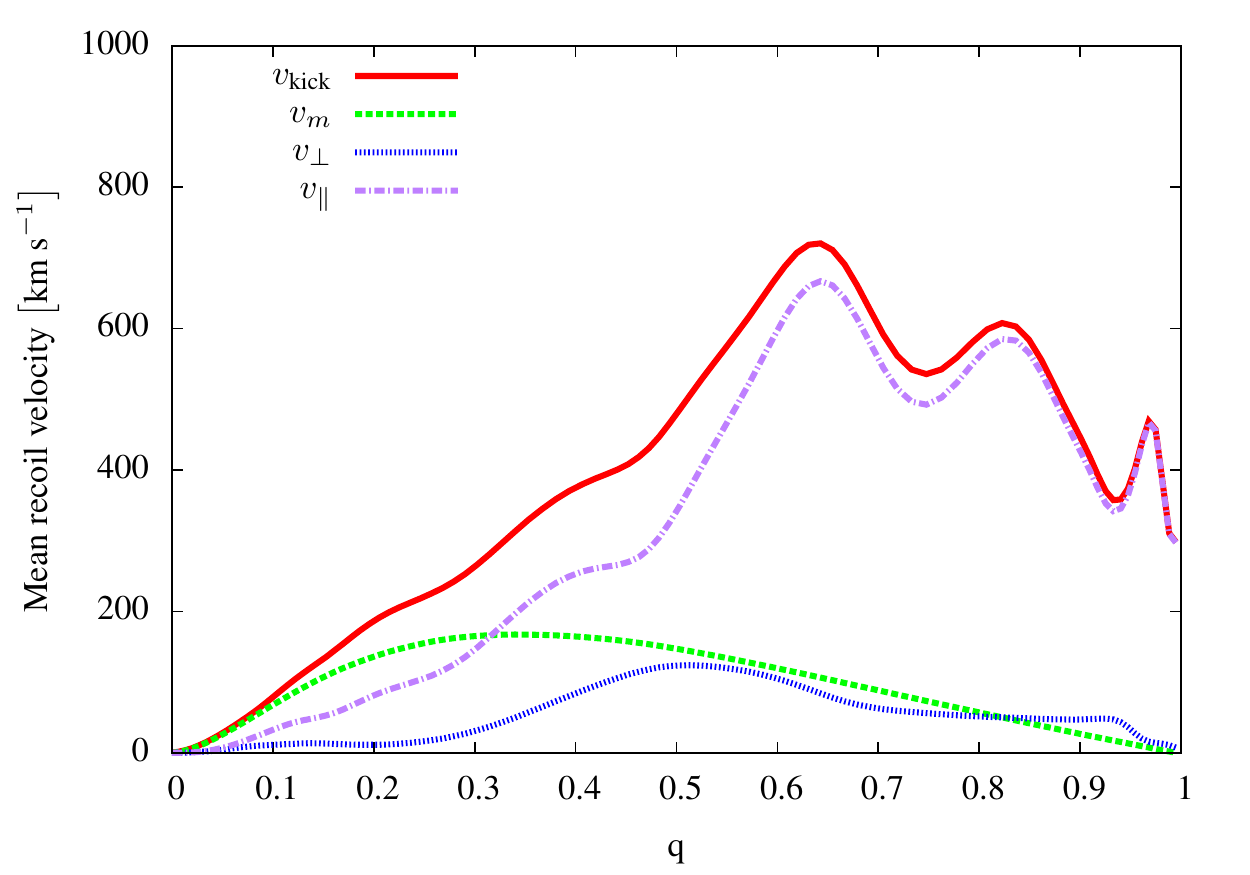}
\caption{Example of how each recoil velocity component depends on $q$. For $v_m$ mass ratio is the only variable, so we have a deterministic formula. Other components and the overall kick velocity depend also on spin magnitudes and orientations, in this case we plot the mean value. As we can see, the major component is almost always $v_\parallel$, others only play a role for low $q$. This plot was done for MS1 mass - spin relationship (e. g. Section \ref{Spin magnitude})}\label{averagekickasafunctionofq}
\end{figure}

The magnitude of GW kick velocity depends on two asymmetries: mass inequality and spin vector asymmetry. These asymmetries are described by the following parameters:
\begin{itemize}
\item
Mass inequality: for two components of a binary we will denote their masses by $m_1$ and $m_2$, with a convention $m_2\leqslant m_1$. A component of the kick related to this asymmetry depends only on the mass ratio $q=\frac{m_2}{m_1}$. For the simplicity of formulas an additional parameter $\eta$, called symmetric mass ratio, is also introduced, defined as:
\begin{equation}\label{eta}
\eta=\frac{m_1m_2}{(m_1+m_2)^2}=\frac{q}{(1+q)^2}
\end{equation}
In our range of interest $q\in[0,1]$ equation \ref{eta} describes a strictly increasing function of $q$ reaching it's maximum value of $\eta=0.25$ for $q=1$.
\item
Spin asymmetry: Each binary component has it's own spin vector $\boldsymbol{J_{\text{BH}}}$. For convenience a dimensionless spin vector is introduced, defined as:
\begin{equation}
\boldsymbol{a_{\text{BH}}}=\frac{c\boldsymbol{J_{\text{BH}}}}{GM_{\text{BH}}^2}
\end{equation}
Physical constraints guarantee that such a vector will have magnitude between $0$ and $1$ (\citet{Angularmomentumlimit}). Unless otherwise specified, in this paper by term spin we will always be referring to this dimensionless vector (in most cases to it's magnitude). Spin related kick component depends on seven parameters: $q$, $a_1$, $\theta_1$, $\varphi_1$, $a_2$, $\theta_2$, $\varphi_2$, where for $i$-th component of a binary $a_{i}$ represents the magnitude, and $\theta_i$, $\varphi_i$ the orientation in a spherical coordinate system, of the vector $\boldsymbol{a_i}$. A spherical coordinate system considered here is centered around $\boldsymbol{e_3}$, direction parallel to orbital angular momentum, with $\varphi=0$ for $\boldsymbol{e_1}=\frac{\boldsymbol{v_m}}{\left|\boldsymbol{v_m}\right|}$ (e. g. Fig. \ref{Kick components drawing}).
\end{itemize}
A common method for calculating kick velocities uses fitting formulas based on post-Newtonian theory and calibrated to simulations. We chose to adapt a fit proposed in \citet{Kickformulas}. It is worth mentioning that recently a more careful method has been proposed in \citet{Gerosawaveformintegration}, which relies on integrating the momentum flux in the gravitational wave. We intend to attempt using this method in later stages of this project.\\
The idea behind recoil velocity formulas derived in \citet{Kickformulas} is best illustrated by Fig. \ref{Kick components drawing}. Formulas are fit for three components: $\boldsymbol{v_m}$ resulting from mass inequality (parallel to the line between BHs just before merger) and $\boldsymbol{v_{\perp}}$, $\boldsymbol{v_\parallel}$, being two projections of a spin asymmetry contribution:
\begin{gather}\label{kickcomponentsformulas}
v_m=A\eta^2\sqrt{1-4\eta}(1+B\eta)\\
v_{\perp}=\frac{H\eta^2}{1+q}\left(a_1\cos\theta_1-qa_2\cos\theta_2\right)\\
v_\parallel=\frac{K\eta^3}{1+q}\left(qa_2\sin\theta_2\cos(\varphi_2-\phi_2)-a_1\sin\theta_1\cos(\varphi_1-\phi_1)\right)
\end{gather} 
$A$, $B$, $H$, $K$ are constants fit in \citet{Kickformulas}. $\phi_1$ and $\phi_2$ are functions of $q$ (constants for a given mass ratio), however they only introduce a rotation in $\varphi$, where we always assume uniform distribution, therefore they can be dropped ($\phi_1=\phi_2=0$) without affecting the results.\\
From Fig. \ref{Kick components drawing} we can conclude that the eventual formula for gravitational recoil should be:
\begin{gather}
\boldsymbol{v_{\text{kick}}}=v_m\boldsymbol{e_1}+v_\perp(\cos\xi\boldsymbol{e_1}+\sin\xi\boldsymbol{e_2})+v_\parallel\boldsymbol{e_3}\\
v_{\text{kick}}=\sqrt{v_m^2+v_\perp^2+2v_mv_\perp\cos\xi+v_\parallel^2}
\end{gather}
In theory angle $\xi$ should be arbitrary, but in \citet{Kickformulas} the real picture was simulated with a simplification where $\xi=\text{const}.$ is another value for the fit, and they obtained $\xi=215^\circ\pm5^\circ$. Therefore we will be also assuming $\xi=215^\circ$.\\
In any case the angle $\xi$ doesn't play a crucial role on the kick velocity, it only appears in a term $2v_mv_\perp\cos\xi$, and, as Fig. \ref{averagekickasafunctionofq} reveals, the major component is almost always $v_\parallel$.
\section{Measures of merger product spin}\label{Finalspin and effective spin}
In \citet{Chieff} (previously e. g. \citet{Oldchieff2}; \citet{Oldchieff1} ) a parameter $\chi_{\text{eff}}$ (effective spin) was introduced as a measure of spin misalignment for BH-BH binaries, defined by:
\begin{equation}\label{chieff}
\chi_{\text{eff}}=\frac{m_1a_1\cos\theta_1+m_2a_2\cos\theta_2}{m_1+m_2}
\end{equation}
It represents a weighed mean value of a spin component parallel to the orbital angular momentum, where each component's impact is weighed by it's mass.\\
The parameter $\chi_{\text{eff}}$ is important in a study of BH-BH mergers, because it is the only parameter we can measure that is (weakly) sensitive to gravitational waveform, therefore it is the only insight into the actual spins that can be gained by an observation of gravitational waves. However, this parameter should not be confused with the actual spin of a product of a merger. In our study, we needed to know how fast does the resulting BH (merger product) rotate, and whether it correlates with the magnitude of recoil velocity. Therefore we needed a parameter of a more straightforward physical interpretation: $a_{\text{fin}}$, the dimensionless spin of a resulting BH.\\
In \citet{Finalspin}  a formula for $a_{\text{fin}}$ was derived by extrapolating (under a few additional assumptions) known formulas and fits for simple cases. This study yields a formula:
\begin{equation}\label{afin}
\begin{split}
a_{\text{fin}}&=\frac{1}{(1+q)^2}\sqrt{a_1^2+q^4a_2^2+q^2\cdot l^2+2q^2a_1a_2\cos\alpha+}\\&\overline{+2q\cdot l\left(a_1\cos\theta_1+q^2a_2\cos\theta_2\right)}
\end{split}
\end{equation}             
where $l$ is the magnitude of a vector $\boldsymbol{l}$ parallel to orbital angular momentum, constituting the additional contribution to final spin vector besides $\boldsymbol{a}_1$ and $\boldsymbol{a}_2$, given by:
\begin{equation}\label{vectorl}
\begin{split}
l&=\frac{s_4}{(1+q^2)^2}\left(a_1^2+q^4a_2^2+2q^2a_1a_2\cos\alpha\right)+\\+&\frac{s_5\eta+t_0+2}{1+q^2}\left(a_1\cos\theta_1+q^2a_2\cos\theta_2\right)+2\sqrt{3}+t_2\eta+t_3\eta^2
\end{split}
\end{equation}
and $\alpha$ is the angle between spin vectors $\boldsymbol{a}_1$ and $\boldsymbol{a}_2$, which can be calculated from $\cos\alpha=\cos\theta_1\cos\theta_2+\sin\theta_1\sin\theta_2\cos(\varphi_2-\varphi_1)$. Constants $t_0=-2.686, t_2=-3.454, t_3=2.353, s_4=-0.129, s_5=-0.384$ are values from numerical fits on which the derivation in \citet{Finalspin} was based on.
\section{Assumptions about spin magnitude and orientation}\label{Spin distributions}
\subsection{Magnitude}\label{Spin magnitude}
There is an ongoing study on the subject of what affects the natal spin of a BH and in particular how does it depend on the BH mass (\citet{Oldmassspinrelation}; \citet{Massspinrelation}). To make some comparisons, we consider 3 different models for mass-spin relationship:
\begin{itemize}
\item MS0: In the mass-spin 0 model, we assume that the spin does not depend on mass. For each BH we draw the spin value at random from a uniform distribution from 0 to 1. This model may not be the most realistic, but it is very useful for understanding the fundamental nature of BH-BH mergers and gravitational kicks, since it covers an entire parameter space without introducing any additional constraints.
\item MS1: We analyzed a metallicity-dependent model introduced in \citet{Massspinrelation}. It predicts high spin for low mass BHs and low spin for massive BHs (details in Section \ref{Remarks on MS1}). Technically the formulas are written for $a$ as a function of $M_{CO}$, mass of a carbon-oxygen core of the progenitor, but, as explained in \citet{Massspinrelation}, $M_{CO}$ is a good proxy for the actual BH mass (for the purpose of our study differences are negligible).
\item MS2: Constant function of $a=a_0=\text{const}.$ This allows to analyze a very straightforward subspace of the parameter space, and varying the value of $a_0$ allows to compare kicks for high-spin and low-spin binaries. We choose to adopt intermediate-value for BH natal spin in core collapse in this model: $a_0=0.5$. Similar approach was adopted in the past ($a_0=0$, $a_0=0.5$, $a_0=0.9$, see \citet{Oldmassspinrelation}).
\end{itemize} 
As already mentioned in Section \ref{trackingblackholes} it is important to stress out that we do not use those mass-spin models for post-merger BHs, for which we assume the spin equal $a_{\text{fin}}$ of the previous merger.
\subsection{Additional remarks on MS1}\label{Remarks on MS1}
As explained thoroughly in \citet{Massspinrelation} the spin value associated with a given CO core mass has a non-monotonic dependence on metallicity ($Z$). Functions are fit for 4 fixed $Z$ values, which for the lack of a better model we extrapolate into 4 metallicity ranges:
\begin{itemize}
\item
range A: $Z<0.001$
\item
range B: $0.001\leqslant Z<0.004$
\item
range C: $0.004\leqslant Z<0.01$
\item
range D: $0.01\leqslant Z$
\end{itemize}
For every range the adopted natal spin function has the following form:
\begin{equation}\label{MS1 spin formula}
a(M)=\begin{cases}
0.85 &\text{ if  } M\leqslant M_1\\
c\cdot M+d &\text{  if  } M_1<M<M_2\\
a_{\text{low}} &\text{  if  } M_2\leqslant M
\end{cases}
\end{equation}
where constants $M_1$, $M_2$, $c$, $d$, $a_{\text{low}}$ are fit separately for each $Z$ range. For our study the most important information are values of $a_{\text{low}}$ and $M_2$, which we provide on top of the Table \ref{MS1 details}, together with other constants adopted from \citet{Massspinrelation}. 
\begin{table}
	\centering
	\caption{Constants used in spin formulas for all metallicity ranges in MS1. $a_{\text{low}}$ and $M_2$ presented on top since they are of biggest interest for us. The non-monotonic nature of this model is clearly visible. It is worth to point out that for BHs above $27.7M_{\astrosun}$ in $Z$ range B spin magnitude $a_{\text{low}}=0$ is assumed.}
	\label{MS1 details}
	\begin{tabular}{ccccc}
		\hline
		& $Z$ range A & $Z$ range B & $Z$ range C & $Z$ range D\\
		\hline
		$a_{\text{low}}$ & $0.25$ & $0$ & $0.25$ & $0.13$\\
		$M_2$ & $38.8M_{\astrosun}$ & $27.7M_{\astrosun}$ & $37.8M_{\astrosun}$ & $24.2M_{\astrosun}$\\
		$M_1$ & $32M_{\astrosun}$ & $18M_{\astrosun}$ & $31M_{\astrosun}$ & $16M_{\astrosun}$\\
		$c$ & $-0.088$ & $-0.088$ & $-0.088$ & $-0.088$\\
		$d$ & $3.666$ & $2.434$ & $3.578$ & $2.258$\\
		\hline
	\end{tabular}
\end{table} 
\subsection{Orientation}\label{Spin orientation}
Our study of BH-BH binaries histories revealed, that the number of interactions they've been through can be extremely large. In particular, they experience numerous binary-binary and binary-single exchanges. A dynamical binary, which has experienced at least one strong interaction, exchange, dissolution, and then binary formation is likely to display deviations from original spin orientation. This will nullify any initial alignment. The number of interactions is too large to attempt assessing resulting spin orientation with satisfying precision (which is one of the reasons there is no reliable spin tracking included in MOCCA and fewbody codes), therefore the best approach is to assume isotropic spin distribution and draw them at random.\\
In a primordial case, if no exchange has taken place, the trace of the initial alignment is still apparent, however, they can experience small deviations from less violent interactions, or because of asymmetric ejection of stellar envelope. For example a fly-by of another star/binary can noticeably alter the direction of the orbital spin vector. We simulate this effect by drawing the $\theta$ angle from a Gaussian around the orbital angular momentum (cut off at $4\sigma$) and $\varphi$ from a uniform distribution, where we leave $\sigma$ of the Gaussian as a free parameter to vary between simulations. Unless otherwise stated, we will be assuming $\sigma=15^{\circ}$.
\section{Finding GW kick retention fractions}\label{Finding rates}
GW kick retention fraction of a given model is defined as a percentage of mergers in that model which occurred inside of the cluster and would not be kicked out of the cluster due to gravitational recoil (see Fig. \ref{whatisafraction}), i. e. those cases when kick velocity is lower than cluster escape velocity at the time of escape. We only consider the gravitational kick, we neglect other possible velocity components such as the natal kick and dynamical kicks obtained by the binary during individual interactions. We approximate the cluster escape velocity based on the cluster central potential at merger time $\Phi_C$ (known from MOCCA output), according to the formula:
\begin{equation}
v_{\text{escape}}=\sqrt{-2\Phi_C}
\end{equation}
This approximation is valid because BHs as the most massive stellar objects typically reside close to the center of the cluster. It should be noted, that we analyze GW kick retention fractions for different models of mass-spin relationship and with different $\sigma$, but to derive these we use data from \underline{all} MOCCA simulations, regardless of their initial assumptions. This means that overall GW kick retention fractions presented in Section \ref{Overall GW kick retention fractions} represent an expected value of GW kick retention fractions, rather than actual GW kick retention fraction in a very specific GC case, which may significantly vary from this result, depending on indivudal properties such as cluster density, mass, tidal forces, binary fraction, and many more. In the following sections we also present GW kick retention fractions limited to specific time bins or cluster types. Mergers are not frequent enough to obtain a reliable statistics for a single cluster, so we have to consider GW kick retention fractions averaged over bigger data sets.
\subsection{Overall GW kick retention fractions}\label{Overall GW kick retention fractions}
\begin{figure}
\centering
\includegraphics[width=\columnwidth]{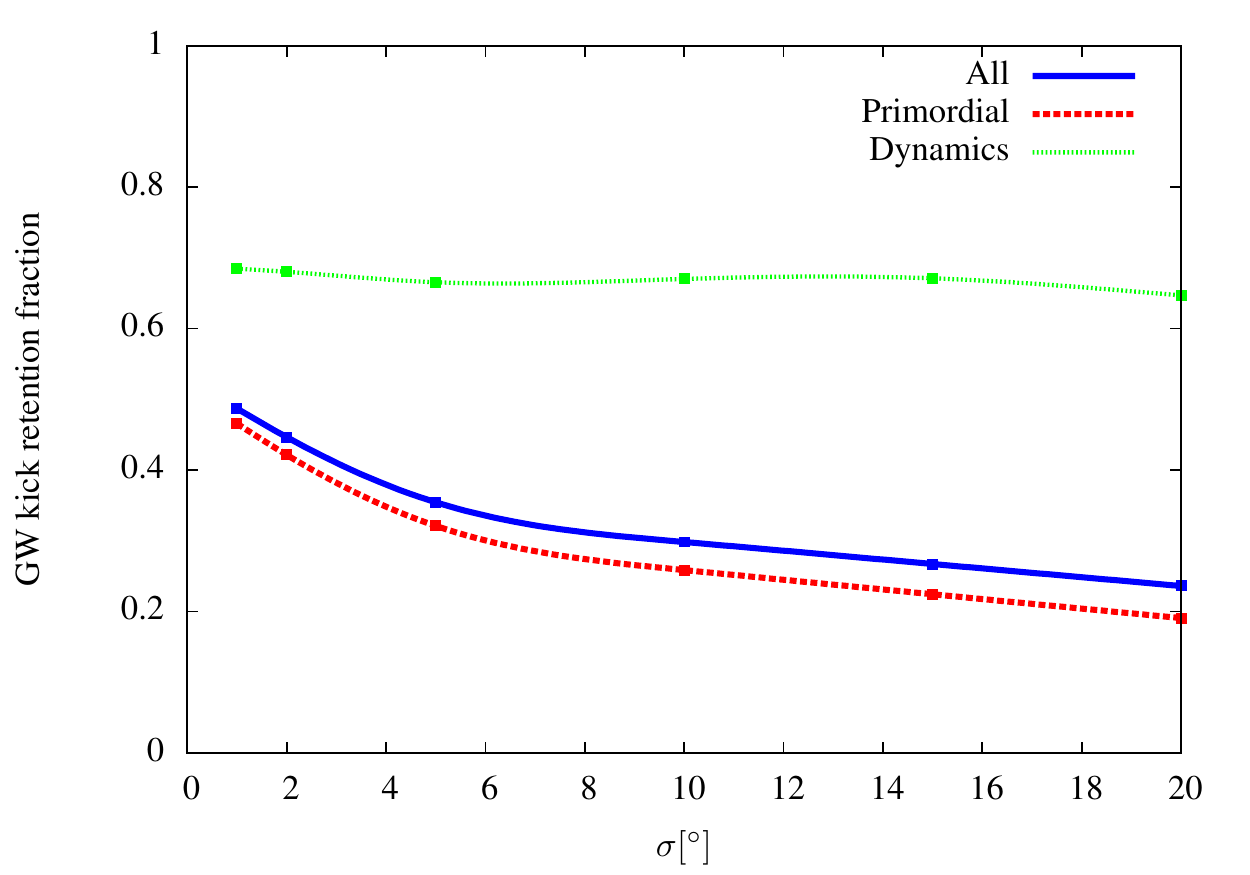}
\includegraphics[width=\columnwidth]{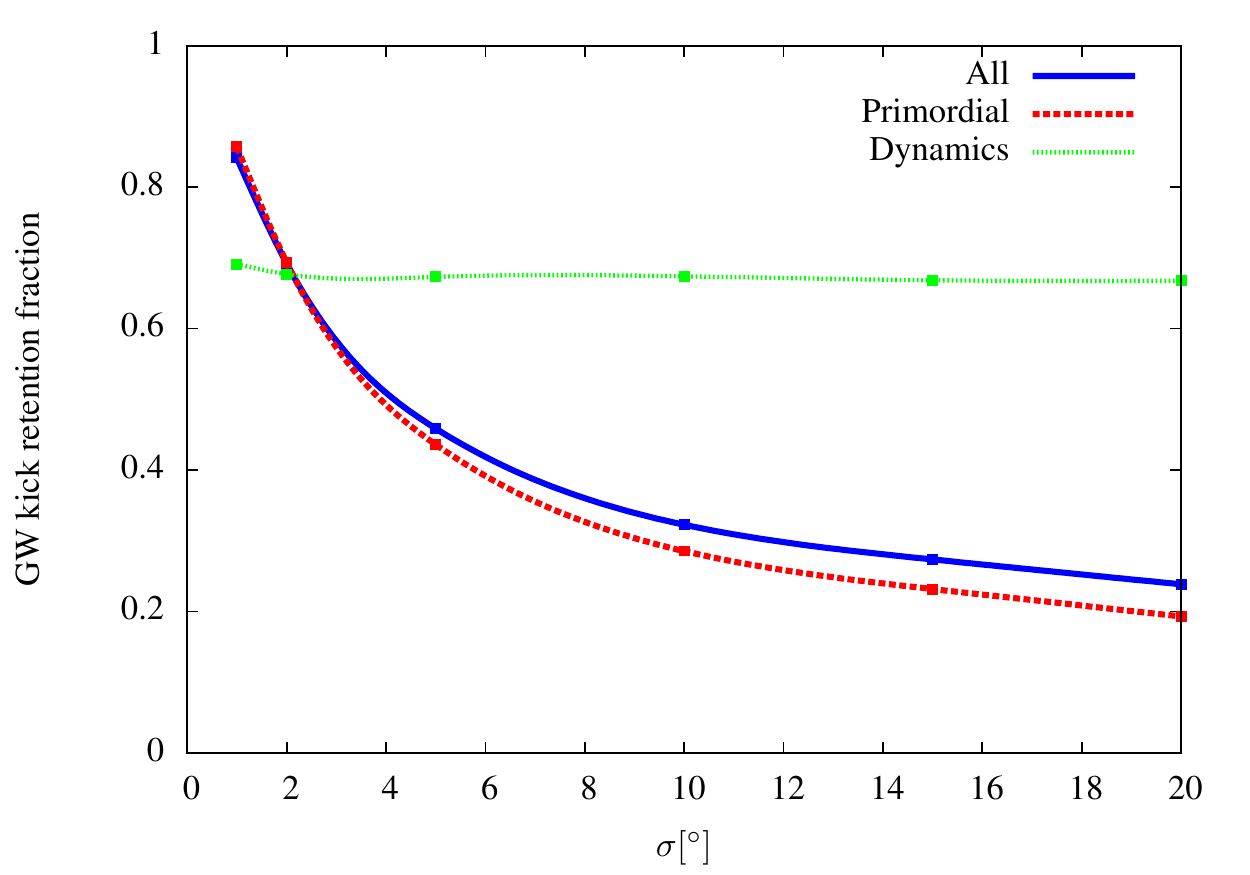}
\includegraphics[width=\columnwidth]{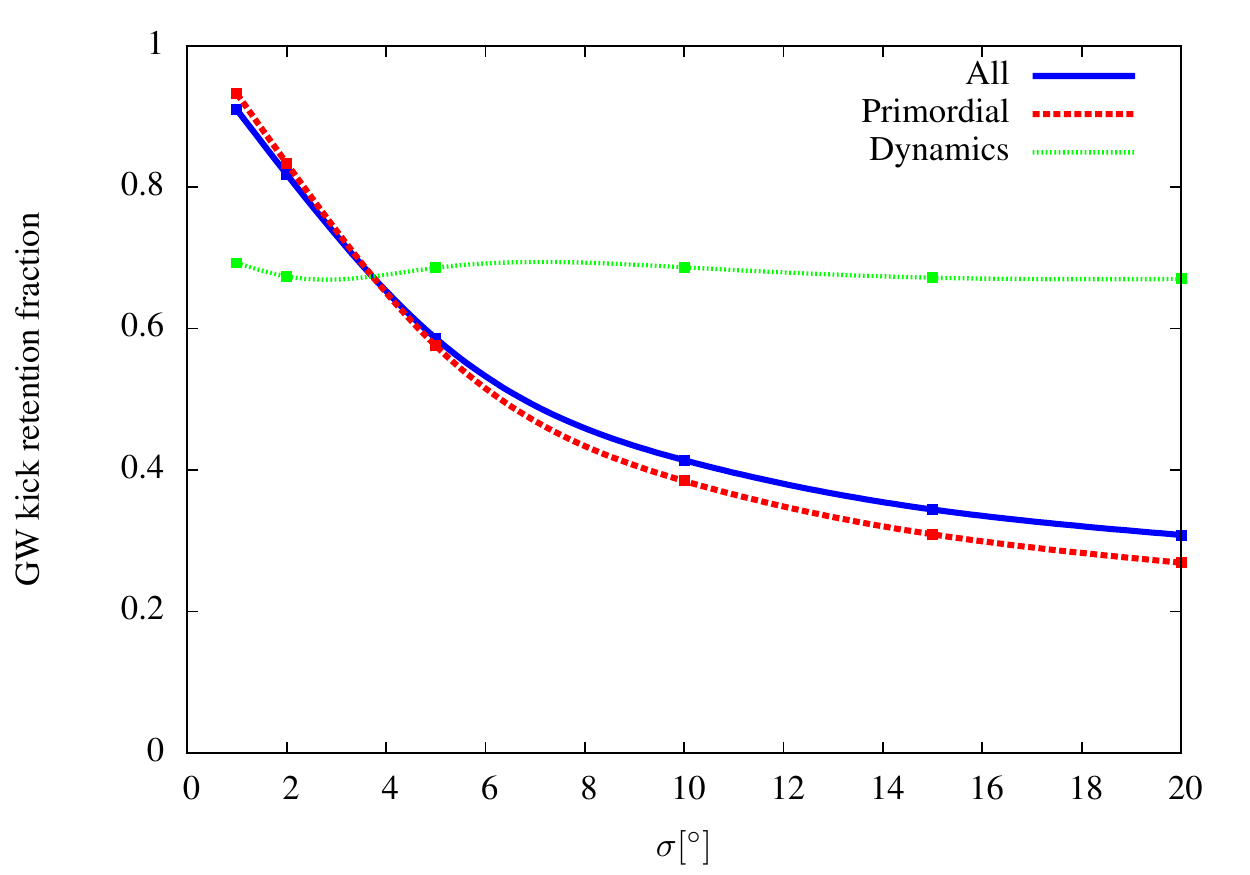}
\caption{GW kick retention fraction as a function of $\sigma$ (deviation of the Gaussian for drawing $\theta$ in primordial case). Top: MS0, middle: MS1, bottom: MS2. Regardless of the model for reasonable values of $\sigma$ ($10^{\circ}-20^{\circ}$) the plot saturates around the value of $0.3$. Therefore the exact value of $\sigma$ assumed doesn't have the crucial effect on the GW kick retention fractions (although it may influence other results). At low $\sigma$ fraction dependence on $\sigma$ is bigger, because $\sigma\sim1^\circ$ impies almost perfect alignment and therefore low kicks ($v_{\parallel}\approx0$, and $v_{\parallel}$ is usually the biggest component, e. g. Fig. \ref{averagekickasafunctionofq}). We also plot GW kick retention fractions for primordial only and dynamical only. The latter displays a small dependence on $\sigma$ even though $\sigma$ is only important for $\theta$ choice in primordial binaries. This is due to the fact that initial spin magnitudes for dynamical mergers in some cases depend on previous primordial mergers, which in turn depend on $\sigma$. However this is a secondary effect with no serious scientific implications.} \label {Theta statistics}
\end{figure}

We obtained a very interesting result that most models lead to very similar overall GW kick retention fractions, i. e. for $\sigma\in[10^{\circ},20^{\circ}]$ for the Gaussian in primordial case (Section \ref{Spin orientation}) fractions don't differ strongly between MS0, MS1 and MS2, and the dependence on $\sigma$ is very small, as we show in Fig. \ref{Theta statistics}. The explanation is that because in MOCCA models escape velocities are around $100\text{ km s}^{-1}$, and average kicks are significantly higher, main impact on the retention is in the general properties of the binary (such as $q$), which come from MOCCA simulation output, and not on exact orientation of spin vectors, which in primordial cases are fairly close to alignment (but not perfectly aligned) in the whole $\sigma\in[10^{\circ},20^{\circ}]$ range. This is supported by typical kick magnitudes shown in Fig. \ref{averagekickasafunctionofq}. 
Nevertheless it can be pointed out that in the range $\sigma\in[10^{\circ},20^{\circ}]$ model MS2 yields significantly higher GW kick retention fractions compared to other two models. This can be explained by our finding that overall GW kick retention fractions depend mostly on GW kick retention fractions for primordial mergers, since they constitute about $90\%$ of evaluated cases. MS1 usually assumes high spin magnitudes for BHs in primordial mergers (because they usually have low mass), and in MS0 half of BHs will also have higher spins than $0.5$ assumed in MS2. Thus MS2 will yield fewer mergers with very high kicks, leading to higher GW kick retention fractions for both primordial and consequently for all mergers.
\subsection{Time dependent GW kick retention fractions}\label{About time dependent rates}
\begin{figure}
\centering
\includegraphics[width=\columnwidth]{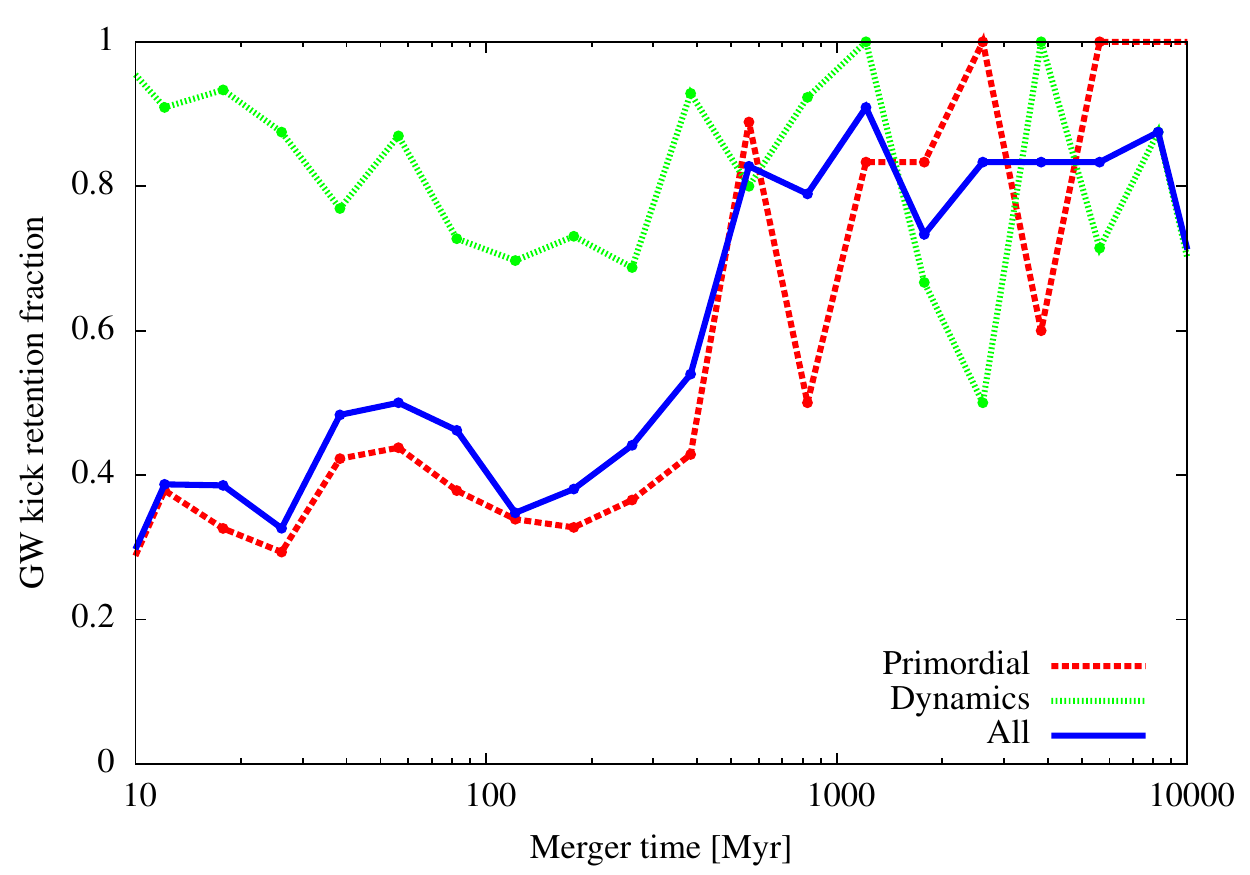}
\includegraphics[width=\columnwidth]{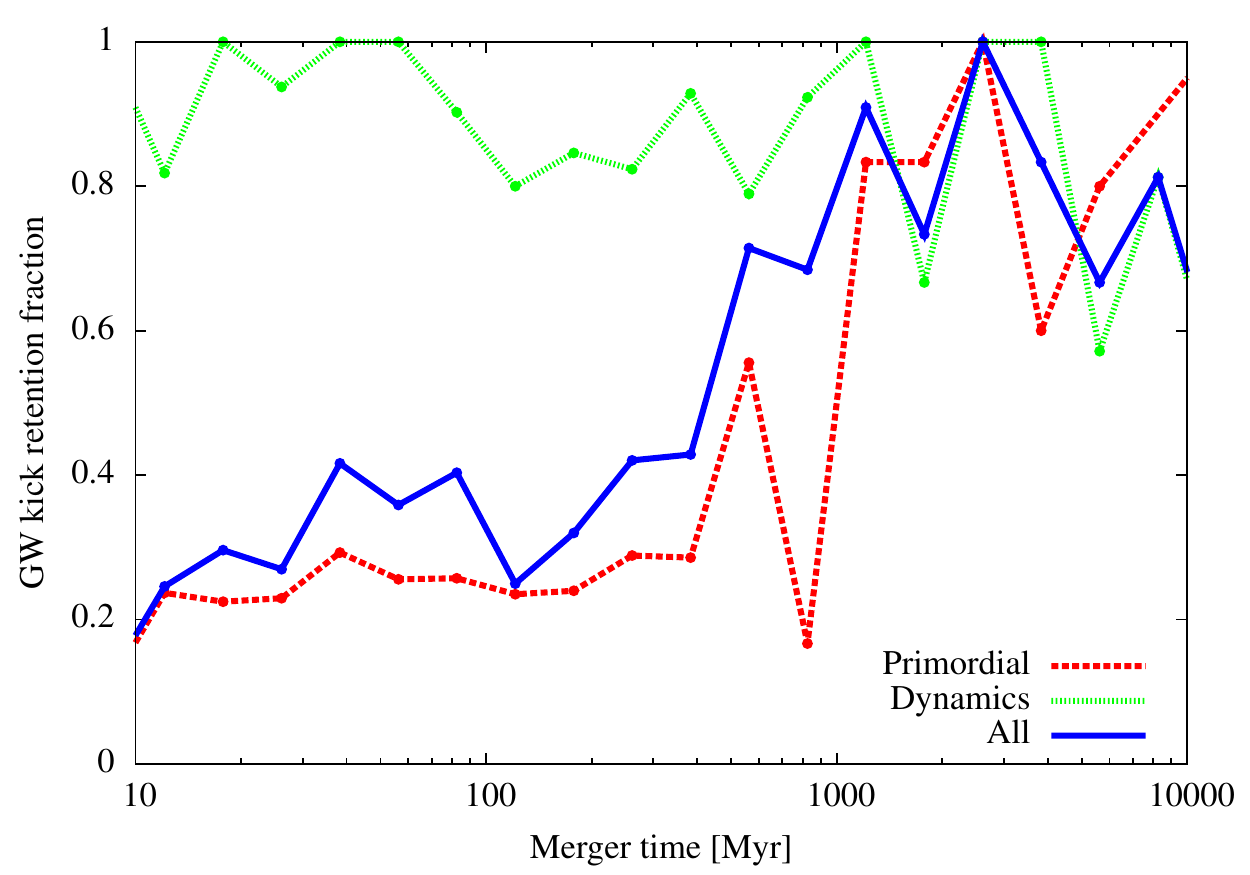}
\includegraphics[width=\columnwidth]{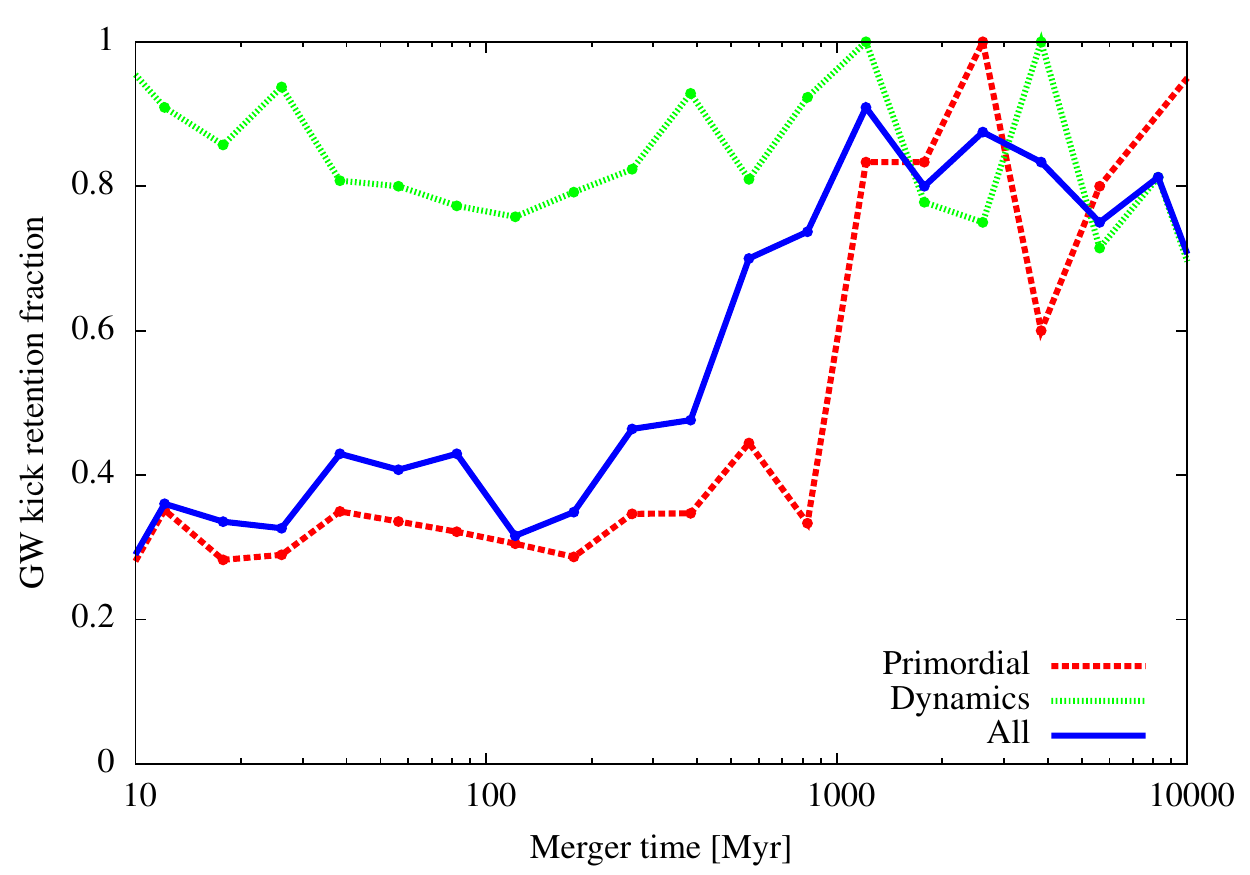}
\caption{Time dependent GW kick retention fractions. Top: MS0, middle: MS1, bottom: MS2. The overall trend is increasing with time, and at the late stages of cluster evolution mergers are almost always retained. Primordial mergers are most likely to be kicked out in MS1 and least likely in MS0.}\label{Plottimedependentrates}
\end{figure}

We also derived time dependent GW kick retention fractions, by binning the logarithm of the merger time, and then applying the standard algorithm for deriving GW kick retention fraction but limited to data from a given bin. The result is presented in Fig. \ref{Plottimedependentrates}.\\
We can observe that primordial merger products are much more likely to be kicked out at the early stages of cluster evolution, and dynamical mergers always have a very high GW kick retention fraction. This directly corresponds with $q$. For primordial mergers it remains very close to $1$ up to time scales of about $10^3$ Myr, and then gradually plummets to about $0.8$, which corresponds to the increase in time dependent GW kick retention fractions. This suggests that low $q$ binaries take longer to merge. For dynamical mergers $q$ is constant with time except for small deviations, which applies to GW kick retention fractions as well.      
\\Figures for MS0, MS1 and MS2 appear similar, but we can see that in MS1 GW kick retention fractions for primordial mergers are significantly lower than corresponding rates in MS0 (these differences only disappear in the last stages of cluster life). This is a representation of the nature of MS1 explained in Section \ref{Remarks on MS1}. Most primordial mergers have low masses, therefore MS1 assumes high spins (usually $0.85$, e. g. equation \ref{MS1 spin formula}), leading to higher kicks than most cases in random MS0. The disappearance of this effect at higher time scales is most likely due to the fact that components of primordial mergers at high time scales have higher masses, indeed we observed that the average value of $m_1+m_2$ is constant at about $20M_{\astrosun}$ but goes up to above $30M_{\astrosun}$ at later times. Therefore MS1 assumes lower spin values in these cases, which nullifies the effect it has at early stages of cluster life.\\
\subsection{Dependence on cluster life}\label{Rates depending on IMBH and subsystem}
\begin{figure}
\centering
\includegraphics[width=\columnwidth]{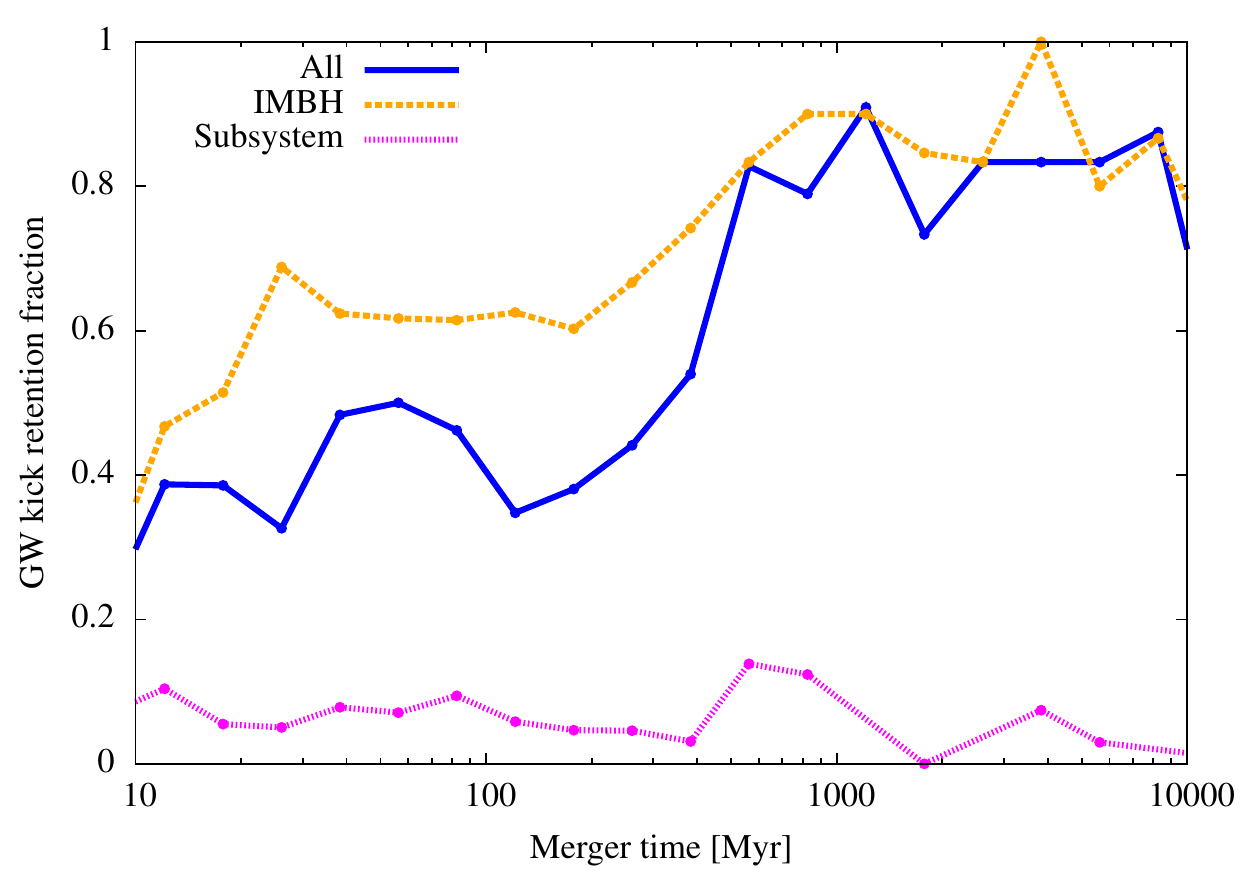}
\includegraphics[width=\columnwidth]{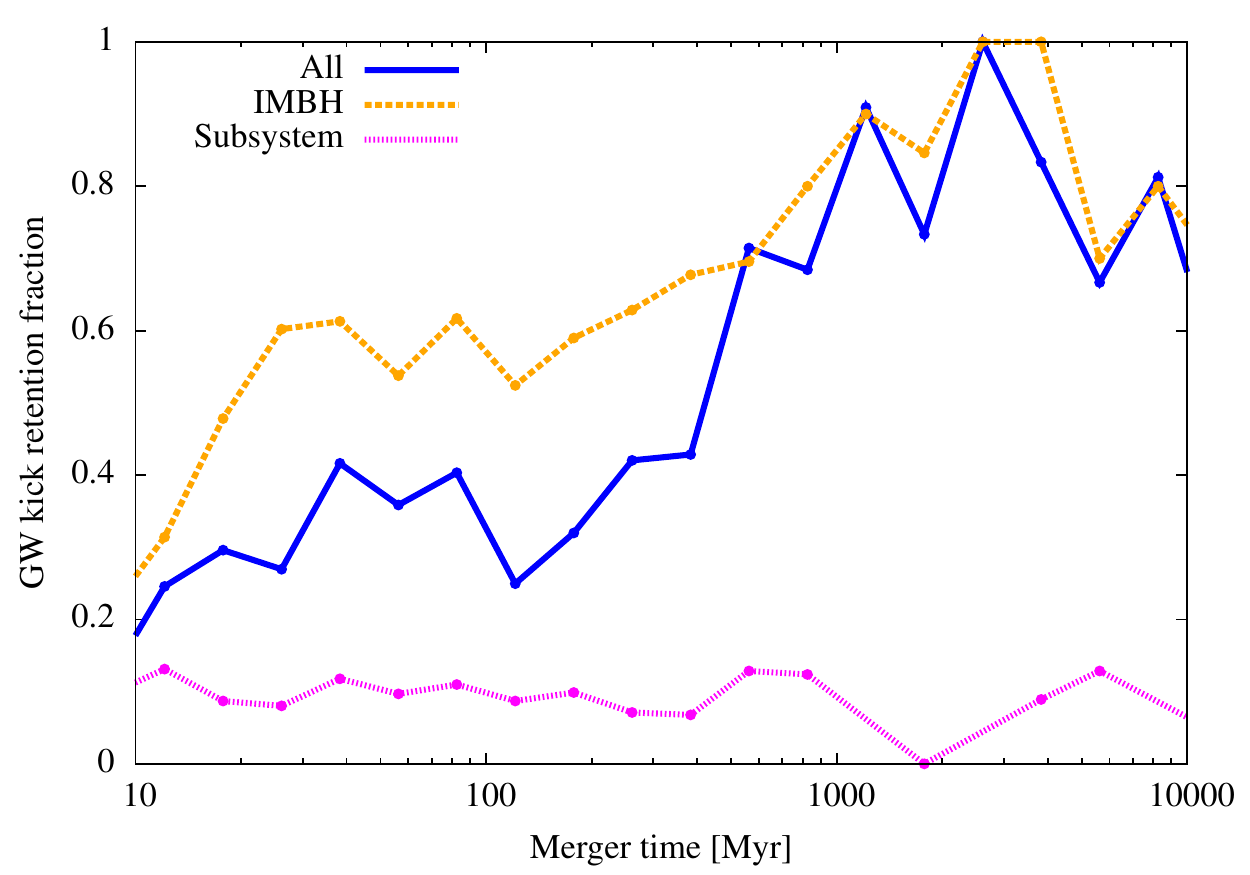}
\includegraphics[width=\columnwidth]{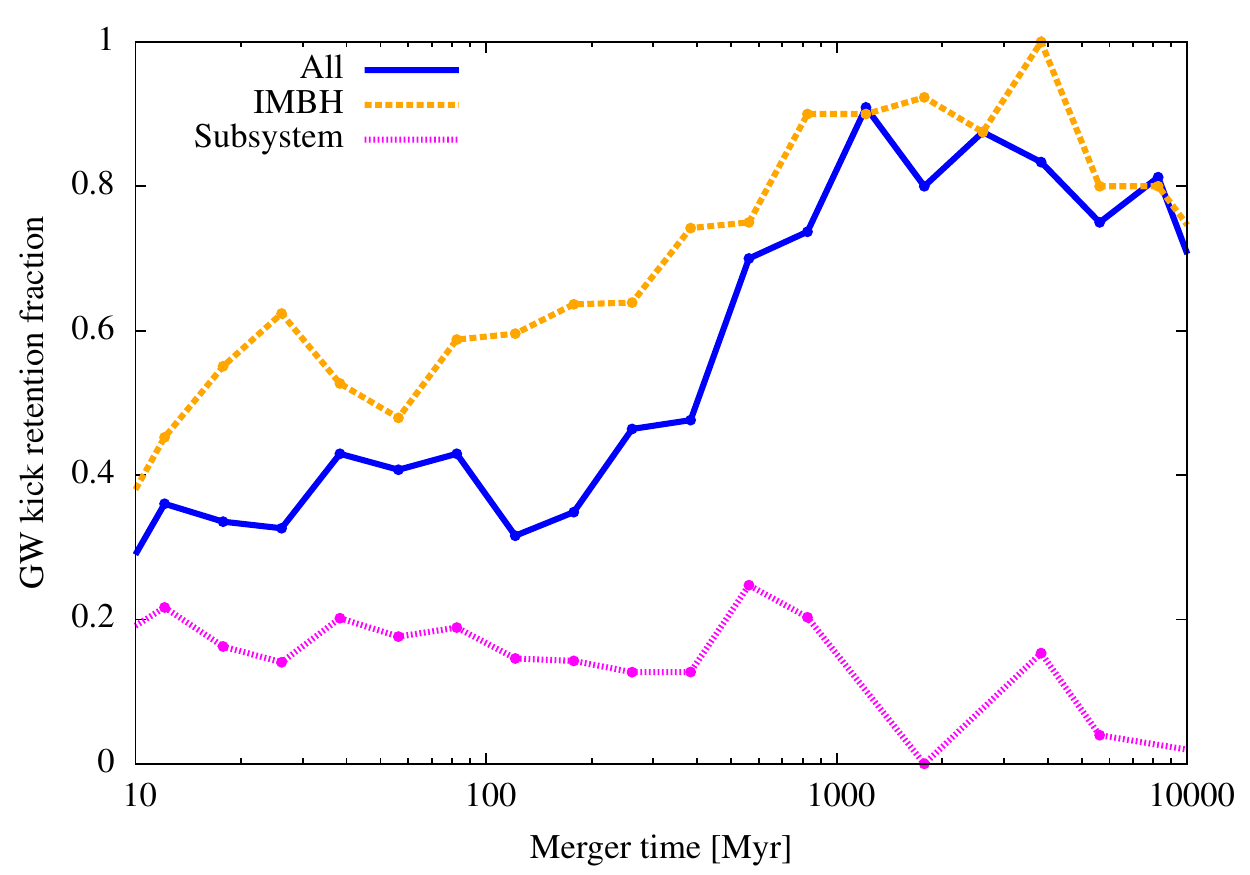}
\caption{Time dependent GW kick retention fractions for models with IMBH or BH subsystem at large time scales. Top: MS0, middle: MS1, bottom: MS2. GCs which form IMBH have highest GW kick retention fractions at all time scales, which is due to low $q$ and partly due to low cluster potential resulting in higher escape velocities. A subsystem has a decreasing impact on GW kick retention fractions.}\label{Plottimedependentratesimbhandsubsystem}
\end{figure}

In order to take a closer look on time dependent GW kick retention fractions, we investigated how they depend on the evolutionary path of the cluster. We considered two possible channels which could have an increasing impact on GW kick retention fractions:
\begin{itemize}
\item
IMBH formation: Not all clusters form an IMBH during their evolution (see \citet{MOCCA2015}), and from MOCCA simulation data we know if a given model has grown an IMBH or not. Existence of an IMBH should increase GW kick retention fractions, because in such cluster most mergers (especially at later times) would be dynamical mergers with the growing IMBH as one of the components, almost bound to be retained, due to very low $q$. A merger between two low mass BHs is unlikely in a cluster with an IMBH. There were $\sim1600$ mergers in clusters with an IMBH.     
\item
Subsystem of BHs: Instead of a single very massive BH, another possibility is that a group of relatively massive BHs would stay in the cluster center.  We call this a subsystem of BHs and we determine these cases by looking if the threshold of $n\geqslant50$ (see \citet{IMBHslow}) is fulfilled at large timescales (12 Gyr, $n$ - number of BHs). There were $\sim600$ mergers in clusters with a subsystem. This proved to be an insufficient sample to be statistically significant, in some time bins there would be just one merger. Therefore we artificially increased the sample by evaluating each of these mergers additional 100 times, drawing new angles for spin orientation, and in case of MS0 new spin magnitudes, in each evaluation (we did it in such a way that these additional artificial mergers would only affect the plot for time dependent fractions in subsystem case in Fig. \ref{Plottimedependentratesimbhandsubsystem}, they were not included in a sample for other studies, otherwise they would distort their results).    
\end{itemize}
We derived separately time dependent GW kick retention fractions in these cases and compared our results with fractions for all clusters, as shown in Fig. \ref{Plottimedependentratesimbhandsubsystem}. Our predictions are confirmed for GCs with an IMBH, but BH subsystems appear to yield lower GW kick retention fractions. It is worth pointing out that clusters with an IMBH have higher than average retention probability already in the early stages of cluster life. Besides low $q$ it is also due to the fact that clusters which form an IMBH usually have significantly higher escape velocities (lower cluster potential) at all timescales, which indeed applies to our data. \\
There are two characteristic scenarios for IMBH formation in GCs (\citet{MOCCA2015}). In the fast scenario a series of consecutive binary mergers and/or collisions occurs in a very dense cluster, leading to a buildup of mass until either a BH with mass above $100M_{\astrosun}$ is formed, or an extremely massive star is formed and subsequently collapses into an IMBH. The slow scenario requires much lower density and a single BH slowly grows in mass around the core collapse time, at time scales of a few Gyr. Plotting merger time against $m_1$ for our data showed that most of the mergers with an IMBH occurred around the time of $\sim10^2-10^3$ Myr, which convinced us that for most of our data the fast scenario applies. This explains why in Fig. \ref{Plottimedependentratesimbhandsubsystem} rates for clusters with an IMBH go up from $\sim0.6$ to $\sim0.8-1$ around the time of $\sim10^3$ Myr - this corresponds to the creation of an IMBH, once a cluster has an IMBH it is very unlikely for merger products to be kicked out, for reasons already explained. \\   
Clusters with a subsystem of BHs yield very low GW kick retention fractions. The most likely explanation is that formation of a subsystem requires simultaneous growth of a big part of BHs in the cluster, so that their potential mergers are likely to have higher than average $q$ (often $q\sim1$, see \citet{IMBHslow}), and thus higher kicks. In such clusters dynamical interactions are very frequent, leading to numerous such mergers. \\
We would like to stress out that the analysis presented here apply to clusters which developed an IMBH or subsystem in the original MOCCA simulation, which did not include GW kicks (see Section \ref{trackingblackholes}). Therefore it applies to clusters which have so far been considered to develop an IMBH or subsystem, but once our work is integrated into MOCCA and new simulations are run we may find out that some of these do not actually form such objects.
\subsection{Dependence on the initial cluster density}\label{Rates depending on rplum}
A dense cluster should be more likely to retain mergers, due to higher escape velocities. We verified it by deriving separate (overall) GW kick retention fractions for three subclasses of cluster models, based on concentration $C=\frac{R_t}{R_h}$ - ratio of tidal radius to half-mass radius (\citet{MOCCA2015}):
\begin{itemize}
\item
Tidally filling: clusters, those which initially fill their Roche lobe. Their $C$ depends on the $W_0$ parameter of the King model (\citet{King}), generally $C\sim10$.
\item
Tidally underfilling, $C=25$
\item
Tidally underfilling, $C=50$ 
\end{itemize}
We present GW kick retention fractions for these three subclasses in Table \ref{Rplum dependent rates}. Dependence on cluster density turns out to be very strong. This stands in agreement with results from \citet{retentionMoody}.\\
From this we can draw the following conclusion: Tidally underfilling clusters yield fairly high GW kick retention fractions and therefore they may support IMBH formation through mergers (fast scenario). Lower density  clusters, on the other hand, have low retention fractions, therefore most likely such clusters might only form IMBHs via the slow scenario, if the cluster collapses before the Hubble time.
\begin{table}
	\centering
	\caption{Overall GW kick retention fractions in three density subclasses of our simulation data. Clearly cluster density plays a very important role on retaining merger products. }
	\label{Rplum dependent rates}
	\begin{tabular}{cccc}
		\hline
		& \makecell{Tidally\\ filling\\ ($C\sim10$)} & \makecell{Tidally\\ underfilling\\ $C=25$} & \makecell{Tidally\\ underfilling\\ $C=50$}\\ 
		\hline
		MS0 & $6.35\%$ & $24.34\%$ & $40.08\%$\\
		MS1 & $7.99\%$ & $25.53\%$ & $39.54\%$ \\
		MS2 & $12.69\%$ & $34.52\%$ & $46.84\%$ \\
		\hline
	\end{tabular}
\end{table} 
\section{Final spin vs effective spin}\label{Finalspin vs effective spin}
\begin{figure}
\centering
\includegraphics[width=\columnwidth]{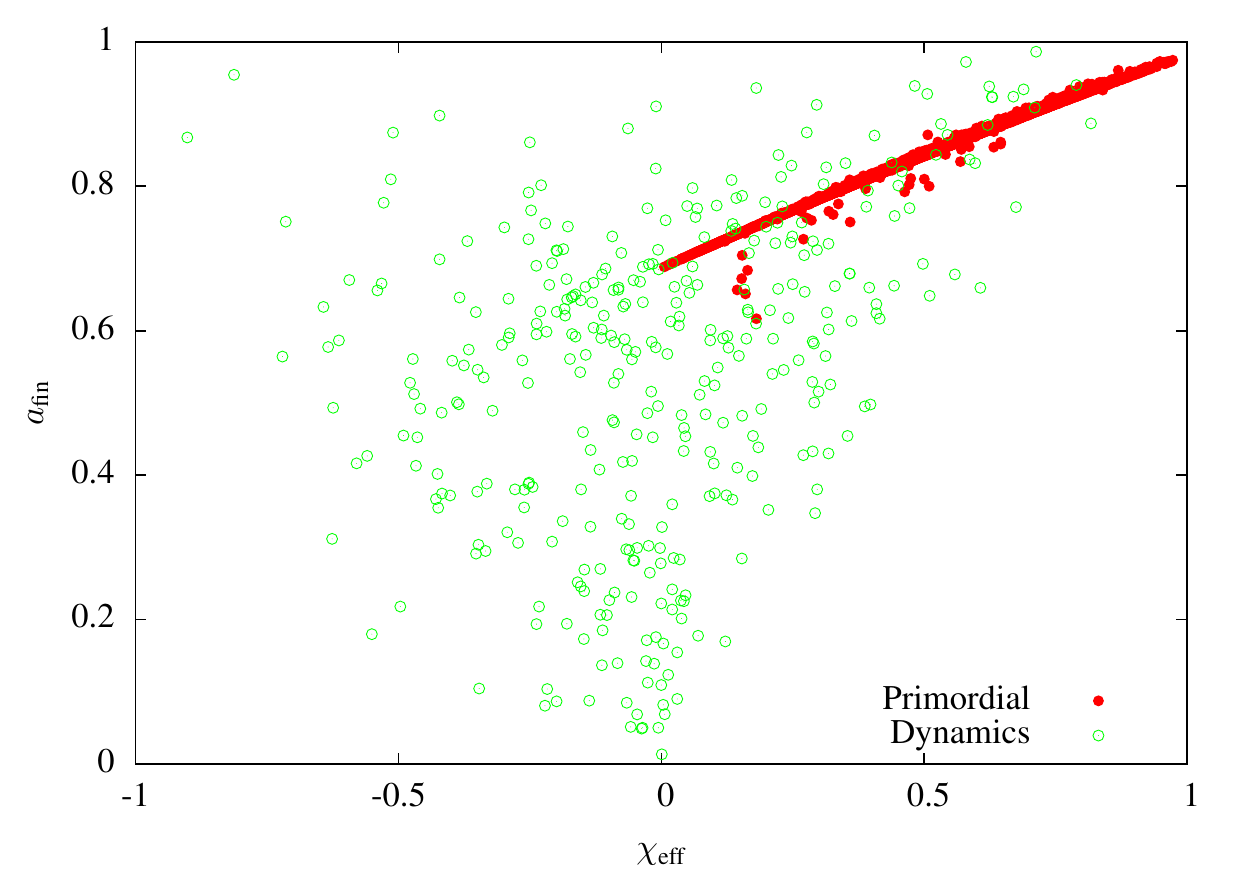}   
\includegraphics[width=\columnwidth]{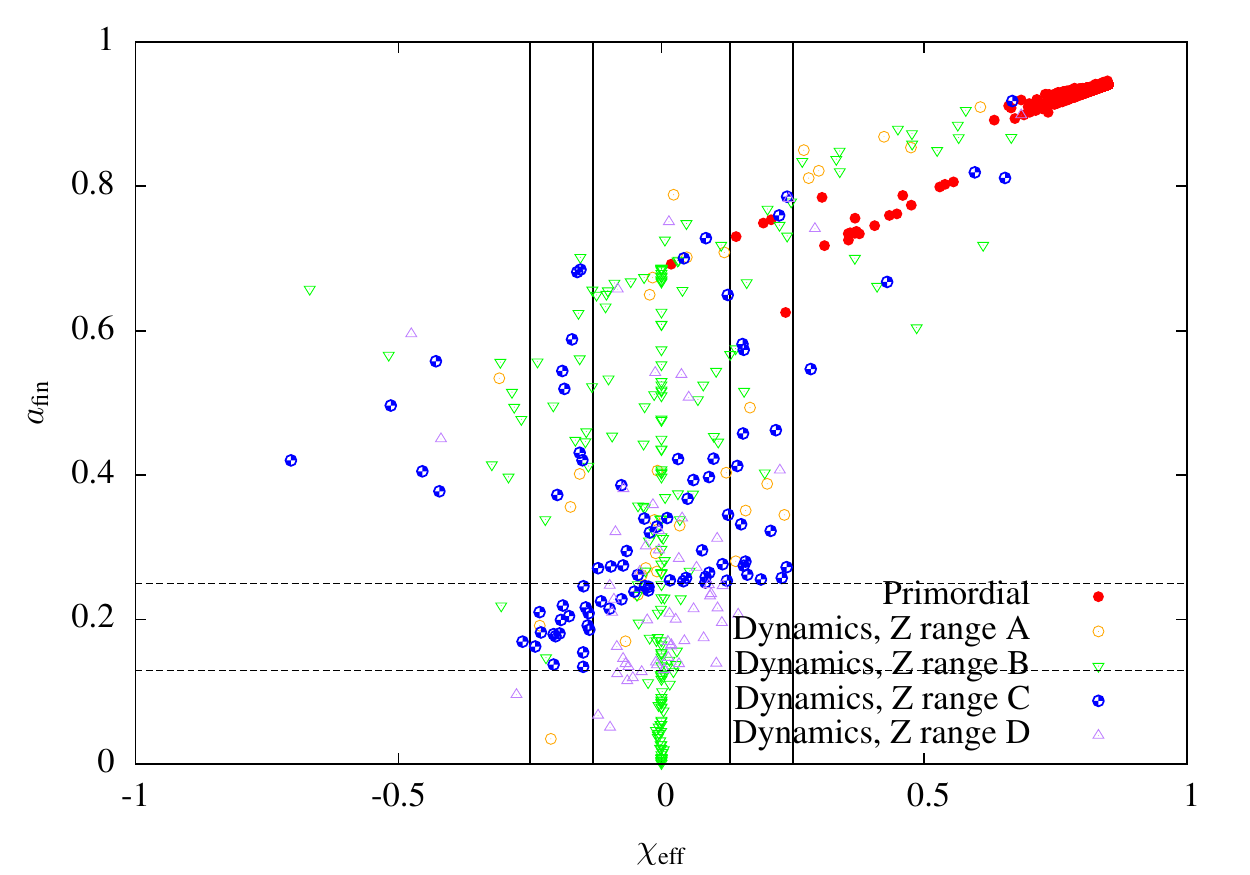}   
\caption{Relation between $\chi_{\text{eff}}$ and $a_{\text{fin}}$, top MS0, bottom MS1 (for MS2 the plot would not be interesting, since in MS2 every BH (except for post-merger BHs) has the same spin). For primordial mergers there is a clear quasi-linear relationship between $\chi_{\text{eff}}$ and $a_{\text{fin}}$. Points deviating from this relationship correspond to $q\lesssim0.8$. For dynamical mergers constraints introduced by MS1 model lead to a very interesting "cross" structure, related to the non-continuous dependence on metallicity. On the bottom picture vertical lines are at $-0.25$, $-0.13$, $0.13$ and $0.25$, horizontal lines at $0.13$ and $0.25$, they indicate areas of $\left|\chi_{\text{eff}}\right|=a_{\text{low}}$, and $a_{\text{fin}}=a_{\text{low}}$ for Z ranges D ($0.13$) and A and C ($0.25$).}\label{Ploteffectivevsfinal}
\end{figure}

We took a close look on parameters describing the effective spin and actual spin of the merger product: $\chi_{\text{eff}}$ and $a_{\text{fin}}$, introduced in Section \ref{Finalspin and effective spin}. Plotting them with respect to each other revealed curious structures shown in Fig \ref{Ploteffectivevsfinal}. 
\subsection{Primordial}\label{Finalspin vs effective spin primordial}
For the primordial case we can see a clear linear relation between effective spin and final spin. To explain this we need to take a look on the typical properties of a primordial merger. As explained in Section \ref{Approximations primordial and dynamics} due to ordered pairing in \citet{Kroupa} algorithm most primordial mergrs have $q\approx1$, that is $m_1\approx m_2$. From the spin orientation drawing method which we adopted, it is clear that in such binaries spins are close to alignment, therefore $\cos\theta_1\approx\cos\theta_2\approx1$. In this case equation \ref{chieff} and equation \ref{afin} simplify to:
\begin{gather}
\chi_{\text{eff}}\approx\frac{a_1+a_2}{2}\\
a_{\text{fin}}\approx\frac{a_1+a_2+l}{4}
\end{gather} 
Additionally from eqation \ref{vectorl} we get:
\begin{equation}
l\approx\frac{\frac{s_5}{4}+t_0+2}{4}(a_1+a_2)+2\sqrt{3}+\frac{t_2}{4}+\frac{t_3}{64}+\frac{s_4}{4}(a_1^2+a_2^2+2a_1a_2)
\end{equation}
Combining these approximate formulas and plugging constants from \citet{Finalspin} yields:
\begin{equation}\label{Primordial spin quadratic approximation}
a_{\text{fin}}\approx0.66+0.4\chi_{\text{eff}}-0.03\chi_{\text{eff}}^2
\end{equation}
The last term is very small, which is why on the plot the relationship appears as linear. Although the explanation above uses some very rough approximations to simplify a much more complex picture, it proves to work very well.\\
In MS1 a vast majority of data points for primordial binaries are clustered in the area  $\chi_{\text{eff}}\in(0.75,0.85)$, with final spin value again in rough accordance with equation \ref{Primordial spin quadratic approximation}. This is a consequence of high spin for low masses in MS1 (see Section \ref{Remarks on MS1}). Most primordial BH-BH binaries have low masses (for both components), which results in $a_1\approx a_2\approx0.85$. In addition, the value of effective spin is lowered by the $\cos\theta$ factor, but since $\theta$ is drawn from a Gaussian with $\sigma=15^{\circ}$ (see Section \ref{Spin orientation}), for 95\% of cases we have $a\cos\theta\geqslant0.85\cdot\cos(30^{\circ})\approx0.74$. This explains why data points are spread around the area of $\chi_{\text{eff}}\in(0.75,0.85)$ (approximately), but are very scarce outside of that interval. Points deviating from this relationship correspond to $q\lesssim0.8$.
\subsection{Dynamics}
\subsubsection{MS0}\label{SpinsdynamicsMS0}
Based on the knowledge that a majority of dynamical mergers have very low $q$ (e. g. Section \ref{Approximations primordial and dynamics}) for the sake of the following analysis we will approximate their $q$ by assuming $q\approx0$ In this case equation \ref{chieff} and equation \ref{afin} simplify to:
\begin{gather}\label{dynamicscaseapproximation}
\chi_{\text{eff}}=\frac{m_1a_1\cos\theta_1+qm_1a_2\cos\theta_2}{m_1+qm_1}\approx a_1\cos\theta_1\\
a_{\text{fin}}\approx a_1 
\end{gather}
from here immediately:
\begin{equation}\label{Spindynamicsinequality}
\left|\chi_{\text{eff}}\right|\leqslant a_{\text{fin}}
\end{equation}
This suggests that there should be a rather straightforward connection between final spin and maximum effective spin, defined as the effective spin assuming perfect alignment, that is:
\begin{equation}\label{maxchieff}
\chi_{\text{eff,max}}=\frac{m_1a_1+m_2a_2}{m_1+m_2}
\end{equation}
Our study confirms that in the regime of small $q$, which applies in most dynamical cases (most dynamical mergers are mergers with an IMBH), we can use maximum effective spin as a proxy for the final resulting spin: $\chi_{\text{eff,max}}\approx a_{\text{fin}}$ (although it becomes less accurate for higher $q$).\\
And because in the dynamical case we assume isotropic distribution for orientation, and the area of a spherical cap cut out by an angle $\theta$ is proportional to $1-\cos\theta$, one should expect that for every value of $a_{\text{fin}}$ data points would be uniformly distributed around the area defined by equation \ref{Spindynamicsinequality}. This stands in agreement with top part of Fig \ref{Ploteffectivevsfinal}.
\subsubsection{MS1}\label{SpindynamicsMS1}
In MS1 we can observe a peculiar "cross" structure: a vertical bar at $\chi_{\text{eff}}=0$ and two horizontal (or slightly slanted, as elaborated on in \ref{Slanted bars}) bars in the low $a_{\text{fin}}$ area, with some high $a_{\text{fin}}$ outliers. It would be especially clear without introducing tracking of individual BHs (see Section \ref{trackingblackholes}), but it is still apparent. Tracking affects only a small fraction of dynamical mergers, therefore from now on for the purpose of this paragraph we will forget about tracking and only consider dynamical mergers without post-merger BHs. \\   
If $a_1\approx a_{\text{fin}}$, it means that the final spin is approximately equal to the initial spin of the more massive component. But from Section \ref{Remarks on MS1} we know that in MS1 high mass objects are assumed to have spin equal $a_{\text{low}}$ of the appropriate $Z$ range. This is indeed the case, as proven by horizontal lines at $0.13$ and $0.25$ on the bottom part of Fig \ref{Ploteffectivevsfinal}, corresponing to $a_{\text{low}}$ for Z range D ($0.13$) and for Z ranges A and C ($0.25$). For Z range B $a_{\text{low}}=0$ (e. g. Table \ref{MS1 details}). In conclusion, different sections of the discussed cross indeed correspond to different metallicity ranges, and within a given range the effect described in Section \ref{SpinsdynamicsMS0}, resulting from a $q\approx0$ approximation, remains in force. This "cross" structure might be mitigated by developing a more precise model with a continuous dependence on metallicity, but GCs don't have continuous $Z$ anyway, therefore some of the "cross" structure will probably always be visible, if there is any dependence of BH rotation on metalicity and initial stellar mass. A closer look on the plot prompts a few additional questions, which we address in Appendix \ref{crossindetail}.
\section{Conclusions}
We derived overall GW kick retention fractions for BH-BH mergers in models with different assumptions concerning initial spin magnitude and orientation for BHs. We introduced a distinction between primordial and dynamical mergers (Section \ref{Approximations primordial and dynamics}) and derived separate GW kick retention fractions for these classes. We analyzed the dependence of retention probability on merger time by deriving GW kick retention fractions limited to specified time intervals (Section \ref{About time dependent rates}). We analyzed dependence of GW kick retention fractions on cluster life and initial density (Section \ref{Rates depending on IMBH and subsystem}, Section  \ref{Rates depending on rplum}).\\
From this we can draw the following conclusions:      
\begin{itemize}
\item
Due to higher than average escape velocities clusters which form an IMBH usually have very high GW kick retention fractions already in the early stages of their evolution.
\item
Clusters with a subsystem of BHs have very low GW kick retention fractions, because mergers in such systems tend to have higher $q$.
\item
Tidally filling clusters have very low GW kick retention fractions. This is strongly related to low escape velocities. For this reason they are unlikely to form IMBH in a fast scenario, but the slow scenario is possible.
\item
Due to high escape velocities dense tidally underfilling clusters retain a fairly high fraction of mergers, especially in the very dense cases ($C\sim50$), therefore they should be able to form IMBHs in a fast scenario.
\item
Retention probability increases with time: for primordial mergers it is much higher in final stages of cluster life than at $10^1-10^3$ Myr. This is related to the plummet of average $q$ at late stages.
\item
All models studied here yield very similar GW kick retention fractions, with the value of $\sim0.3$. This is due to the fact that other binary properties (primarily $q$) have a much stronger impact on recoil velocities than the spin vectors.

\item
Our results regarding GW kick retention fractions, their dependence on cluster density, age and high retention probability for dynamical mergers involving an IMBH generally agree with previous semi-analytical studies.
\item
Distinction between primordial and dynamical mergers for spin orientation drawing method is important, in particular the value of $\sigma$ for the Gaussian in a primordial case cannot be too small, but should be around $10^{\circ}-20^{\circ}$. Otherwise GW retention fractions turn out to be very high and strongly sensitive to $\sigma$, because low $\sigma$ implies almost perfect alignment and very low kicks.
\item
For $\sigma\in[10^{\circ},20^{\circ}]$ model MS1 yields slightly lower GW kick retention fractions than simpler models known before. This is due to the fact that in primordial mergers BH masses are usually very low, and MS1 assigns them high spin magnitudes.
\item
In studies like this one it is important to track individual BHs, otherwise some conclusions can be simplified and too optimistic (higher GW kick retention fractions). 
\item
Model MS1 introduces some very interesting effects, clustering data points in very specific regions of the parameter space: the "cross" formed on the plot of final spin vs effective spin and a range of non-spinning binaries with a very clear relationship between final spin and recoil velocity. These effects have not been seen before (in MS2 or MS0), since they are caused by a non-continuous spin dependence on metallicity assumed in this model.   
\end{itemize}
In the future studies we are planning the following improvements and extensions to our methods:
\begin{itemize}
\item
Deriving GW kick retention fractions for BH-BH mergers in binary-single and binary-binary interactions.
\item
Considering a broader range of models concerning spin orientation. 
\item
Considering a broader range of values for $\alpha$ and $\lambda$ - constants determining common envelope phase treatment in MOCCA - this would yield various data samples some of which would have much higher percentage of dynamical mergers. 
\item
Investigating the impact of updated assumptions in MOCCA (future data releases, for example a modification of the mass pairing algorithm from \citet{Kroupa}, introduced in \citet{Initialbinarypopulationcorrection}) on GW kick retention fractions.
\item
Improving BH tracking so as to follow an entire history of each BH and assess the impact on spin from all past interactions.
\item
Attempting an alternative method for computing kick magnitudes proposed in \citet{Gerosawaveformintegration}
\item
Eventually incorporating gravitational kicks and resulting escapes into MOCCA code.
\end{itemize}
\section*{Acknowledgements}

We would like to thank Nicolas Copernicus Astronomical Center for hosting the summer internship program, during which this project was carried out. Personally as Jakub Morawski I want to acknowledge Mirek Giersz for advice as an internship supervisor, Abbas Askar for providing files with a selection of BH-BH mergers from MOCCA simulation, which saved hours of digging through output data, and Krzysztof Belczynski for arranging this internship for me and joining for discussion of the results. We would also like to thank Richard O'Shaughnessy for help with resolving our doubts concerning spin orientation drawing method. MG was partially supported by the National Science Center (NCN), Poland, through the grant UMO-2016/23/B/ST9/02732. AA was partially supported by NCN, Poland, through the grants UMO-2016/23/B/ST9/02732 and UMO-2015/17/N/ST9/02573. AA is currently supported by the Carl Tryggers Foundation through the grant CTS 17:113. KB acknowledges support from the Polish National Science Center (NCN) grants: Sonata Bis 2 (DEC-2012/07/E/ST9/01360), LOFT/eXTP (2013/10/M/ST9/00729) and OPUS (2015/19/B/ST9/01099). This work benefited from support by the International Space Science
Institute, Bern, Switzerland,  through its International Team programme
ref. no. 393 {\it The Evolution of Rich Stellar Populations \& BH
Binaries} (2017-18).




\bibliographystyle{mnras}
\bibliography{example}



\appendix

\section{More detailed analysis of the cross-like structure in MS1 final spin vs effective spin plot}\label{crossindetail}

\subsection{What happens in $Z$ range B?}\label{ZrangeBissue}
If the explanation by approximation from equation \ref{dynamicscaseapproximation} were complete, for $Z$ range B we should expect all data points in the vicinity of $\chi_{\text{eff}}=a_{\text{fin}}=0$, instead we can see a vertical bar at $\chi_{\text{eff}}=0$. This phenomenon seems to undermine our explanation and a more detailed analysis is necessary to agree the two. \\
As it turns out, all data points above $a_{\text{fin}}\approx0.2$ constitute outliers representing less than $20\%$ of the data (in Z range B). A closer look on Fig. \ref{Ploteffectivevsfinal} suggests that similar outliers appear also in other metallicity ranges, but they're more scattered, whereas for $Z$ range B most of them lie along the line of $\chi_{\text{eff}}=0$, which leads to a fake impression that these data points have a greater meaning.
\subsection{Why is there a slight increasing trend in the bars of the cross?}\label{Slanted bars}
In Fig \ref{Ploteffectivevsfinal} we can see that for metallicity ranges A, C and D, although their final spin lies along the horizontal line of a fixed value ($0.25$ for A and C, $0.13$ for D), most data points land below this line for $\chi_{\text{eff}}<0$ and above it for $\chi_{\text{eff}}>0$.\\
This effect can be explained by relaxing the assumption of $q=0$ into saying that $q$ is small, so we can ignore every term with higher powers of $q$ in the formula \ref{afin}, but we leave the linear term: $a_{\text{fin}}\approx\frac{1}{(1+q)^2}\sqrt{a_1^2+2qla_1\cos\theta_1}\approx\frac{a_1}{1+2q}\sqrt{1+2q\frac{l}{a_1}\cos\theta_1}$. Since $q$ is small we can use approximations $(1+x)^\alpha\approx1+\alpha x$ for small $x$, which gives:
\begin{equation}
\begin{split}
a_{\text{fin}}&\approx a_1\left(1+q\frac{l}{a_1}\cos\theta_1\right)\left(1-2q\right)\approx a_1-2a_1q +\\&+q\frac{l}{a_1}a_1\cos\theta_1\approx a_1-2a_1q+q\frac{l}{a_1}\chi_{\text{eff}}
\end{split}
\end{equation}               
where the last approximation comes from equation \ref{dynamicscaseapproximation}. We can see a term linear with effective spin (although $l$ also has non-trivial dependence on $a_1$, $a_2$, $q$ etc., in this case we can safely say it's constant due to the dominant term of $2\sqrt{3}$, and for the bars of the cross $a_1$ is constant as well), which explains the observed trend. There is also an extra term $-2qa_1$: it should move the plot slightly lower, but apparently this effect is too small to be noticed, or perhaps it's mitigated by higher order effects.
\subsection{Outliers in effective spin - final spin plot in MS1}\label{Outliers}
\begin{figure}
\centering
\includegraphics[width=\columnwidth]{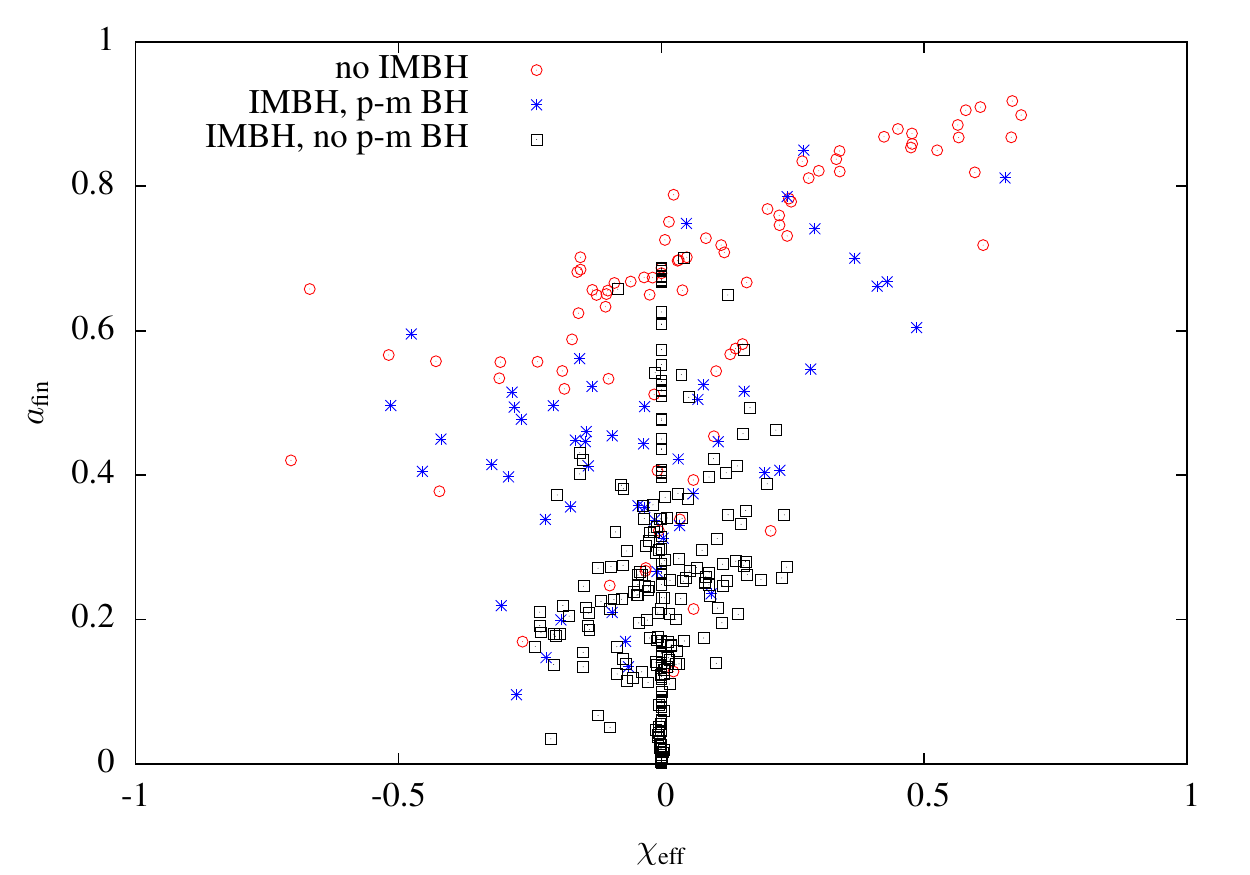}
\caption{Distinction between dynamical interactions involving an IMBH with spins calculated according to MS1 without post-merger BH (here p-m BH), with IMBH and p-m BH, and others (there is no use from filtering out cases with p-m BH and no IMBH as a separate category, because there are very few such cases). Clearly the last two categories are responsible for almost all of the high $a_{\text{fin}}$ outliers.}\label{Crossoutliersexplained}
\end{figure}
As mentioned in Section \ref{Approximations primordial and dynamics}, about $77\%$ of dynamical mergers in our data were mergers in which one of the components was an IMBH ($m_1>100M_{\astrosun}$). The remaining $23\%$ would have much higher $q$ and therefore they lie outside of the regime where approximation $q\approx0$ is applicable. As we show in Fig \ref{Crossoutliersexplained}, these cases are responsible for most of the high spin outliers on the $a_{\text{fin}}$ - $\chi_{\text{eff}}$ plot (data points outside of the cross-like structure). There rest of outliers neatly correspond with the effect of tracking, as almost all of them represent binaries with a post-merger component. 


\bsp	
\label{lastpage}
\end{document}